\documentclass[iop,twocolappendix]{emulateapj}
\usepackage{apjfonts}
\usepackage{amsmath,amssymb}
\renewcommand{\vec}[1]{\mathbf{#1}}
\newcommand{\diffp}[2]{\frac{\partial #1}{\partial #2}}
\newcommand{\sgn}{\mathrm{sgn}\,}
\shorttitle{Alpha-Particle Drift-Anisotropy Instabilities}
\shortauthors{D.~Verscharen et al.}

\begin{document}

\title{Instabilities Driven by the Drift and Temperature Anisotropy of Alpha Particles in the Solar Wind}


\author{Daniel Verscharen, Sofiane Bourouaine, and Benjamin D.~G.~Chandran \altaffilmark{1}}
\affil{Space Science Center, University of New Hampshire, Durham, NH 03824, USA; daniel.verscharen@unh.edu, s.bourouaine@unh.edu, benjamin.chandran@unh.edu}
\altaffiltext{1}{Also at Department of Physics, University of New Hampshire, Durham, NH 03824, USA}

\journalinfo{The Astrophysical Journal, 773:163 (13pp), 2013 August 20}
\submitted{Received 2013 April 5; accepted 2013 July 6; published 2013 August 6}

\begin{abstract}
We investigate the conditions under which parallel-propagating
Alfv\'en/ion-cyclotron (A/IC) waves and fast-magnetosonic/whistler
(FM/W) waves are driven unstable by the differential flow and
temperature anisotropy of alpha particles in the solar wind. We focus
on the limit in which $w_{\parallel \alpha} \gtrsim 0.25 v_{\mathrm A}$, where
$w_{\parallel \alpha} $ is the parallel alpha-particle thermal speed
and $v_{\mathrm A}$ is the Alfv\'en speed. We derive analytic expressions
for the instability thresholds of these waves, which show, e.g., how
the minimum unstable alpha-particle beam speed depends upon~$w_{\parallel \alpha
 }/v_{\mathrm A}$, the degree of alpha-particle temperature
anisotropy, and the alpha-to-proton temperature ratio.  We
validate our analytical results using numerical solutions to the full
hot-plasma dispersion relation. Consistent with previous work, we find
that temperature anisotropy allows A/IC waves and FM/W waves to become
unstable at significantly lower values of the alpha-particle beam
speed~$U_\alpha$ than in the isotropic-temperature case. Likewise,
differential flow lowers the minimum temperature anisotropy needed to
excite A/IC or FM/W waves relative to the case in which $U_\alpha =
0$. We discuss the relevance of our results to alpha particles in the
solar wind near 1~AU.
\end{abstract}


\section{Introduction}

The solar wind is a magnetized plasma outflow originating from the
solar corona and filling interplanetary space. It consists of protons,
electrons, alpha particles, and minor ions. The alpha particles
comprise $\sim 15$\% of the mass density of the fast solar
wind~\citep{bame77}.  The Coulomb collision timescale for ions in the
fast wind is typically much larger than the wind's approximate travel
time from the Sun~$r/U$, where $r$ is heliocentric distance and $U$ is
the proton outflow speed~\citep{kasper08}. Because collisions are
weak, the expansion and heating of the solar wind cause the plasma to
develop non-Maxwellian features, including temperature anisotropies
and ion beams~\citep{marsch82b,goldstein00,reisenfeld01}.  Such
non-Maxwellian features provide a source of free energy that can drive
plasma instabilities whose growth timescales are much shorter than
$r/U$ \citep{hasegawa72,montgomery76,gomberoff91,daughton98,podesta11}. If the plasma evolves beyond the threshold of such an
instability, the instability grows and the resulting electromagnetic
fluctuations can interact with particles to reduce the source of free
energy that drives the instability \citep{kaghashvili04,lu09}.  For example, it has been argued
that instabilities in the solar wind limit the degree of proton and
alpha-particle temperature anisotropies
\citep{kasper02,hellinger06,matteini07,bale09,maruca12} and the speeds of proton
and alpha-particle beams \citep{gary00,hellinger11}.

The thresholds of the instabilities that are driven by alpha-particle
beams depend  on the value of $w_{\parallel \alpha}/v_{\mathrm A}$, where
$w_{\parallel \alpha}$ is the parallel thermal speed of the alpha
particles,
\begin{equation}
v_{\mathrm A} = \frac{B_0}{\sqrt{4\pi n_{\mathrm p} m_{\mathrm p}}}
\end{equation} 
is the proton Alfv\'en speed, ${\bf B}_0$ is the background magnetic
field, and $n_{\mathrm p}$ and $m_{\mathrm p}$ are the proton number density
and mass. We first summarize some previous results pertaining to the
isotropic-temperature case, in which
$T_{\perp \alpha}=T_{\parallel \alpha}$ and $T_{\perp \mathrm p} = T_{\parallel \mathrm
  p}$, where $T_{\perp \alpha}$ ($T_{\parallel \alpha}$) and
$T_{\perp \mathrm p}$ ($T_{\parallel \mathrm p}$) are the perpendicular
(parallel) alpha-particle and proton temperatures, respectively. When
$w_{\parallel \alpha} \ll v_{\mathrm A}$, parallel-propagating
Alfv\'en/ion-cyclotron (A/IC) waves are stable. We use the phrase
``parallel-propagating'' to describe waves with ${\bf k} \times {\bf
  B}_0 = 0$, where ${\bf k}$ is the wavevector. On the other hand,
oblique A/IC waves (with ${\bf k}\times {\bf B}_0 \neq 0$) become
unstable when $U_\alpha \gtrsim v_{\mathrm A}$, where ${\bf U}_\alpha$ is
the average alpha-particle velocity as measured in the proton
frame~\citep{gary00,verscharen13}. In the opposite limit---i.e.,~ $w_{\parallel \alpha} \gtrsim v_{\mathrm A}$---the parallel-propagating A/IC wave
becomes unstable when~$U_\alpha \gtrsim v_{\mathrm
  A}$~\citep{verscharen13a}. The parallel-propagating
fast-magnetosonic/whistler (FM/W) wave is unstable when $U_\alpha
\gtrsim 1.5 v_{\mathrm A}$ provided $w_{\parallel \alpha}\gtrsim 0.3
v_{\mathrm A}$, but at smaller $w_{\parallel \alpha}/v_{\mathrm A}$ it becomes
increasingly difficult to excite this
instability~\citep{gary00,verscharen13}. The dependence of the FM/W
instability threshold on $w_{\parallel \alpha}/v_{\mathrm A}$, however, has
not been described quantitatively in previous work, to the best of our
knowledge.

The above results apply in the case of isotropic temperatures, but in
general $T_{\perp \alpha} \neq T_{\parallel \alpha}$ and $T_{\perp \mathrm
  p} \neq T_{\parallel \mathrm p}$ in the solar
wind~\citep{marsch82,marsch82b}.  Previous work has shown that
temperature anisotropy can significantly reduce the $U_\alpha$
threshold of both the A/IC and FM/W instabilities when $w_{\parallel \alpha} \gtrsim v_{\mathrm
  A}$~\citep{araneda02,gary03,hellinger03}. In principle, the growth or damping rates of A/IC and FM/W waves depend upon other plasma parameters as well. 

In this work, we derive analytic expressions for the instability thresholds of parallel-propagating A/IC and FM/W waves that show how these thresholds depend upon $w_{\parallel \alpha}/v_{\mathrm A}$,
$U_{\alpha}/v_{\mathrm A}$, $T_{\perp \alpha}/T_{\parallel \alpha}$,
 $T_{\parallel \alpha}/ T_{\parallel \mathrm p}$, and $n_{\alpha}/n_{\mathrm p}$, where $n_{\alpha}$ is the
alpha-particle number density.
 For simplicity, we
keep $T_{\perp \mathrm p} = T_{\parallel \mathrm p}$ throughout our analysis. We also focus on the limit in which $w_{\parallel \alpha} \gtrsim 0.25
v_{\mathrm A}$. At smaller $w_{\parallel\alpha}$, our theory is not applicable because the approximate dispersion relations do not reproduce the resonant wavenumber range to the necessary degree of accuracy. One of our motivations for carrying out this calculation
is to enable more detailed comparisons between these instability
thresholds and spacecraft measurements. Such comparisons will be
important for furthering our understanding of the extent to which
instabilities limit differential flow and alpha-particle temperature
anisotropy in the solar wind. A second motivation for our carrying out
an analytic calculation is that it illustrates the physical processes
that determine the instability criteria, thereby offering additional
insights into these instabilities.

The remainder of this paper is organized as follows. In
Section~\ref{sec:framework}, we briefly review the linear theory of
resonant wave--particle interactions.  In Section~\ref{sec:criteria}, we
apply this theory to derive the instability thresholds for A/IC waves
and FM/W waves and compare these thresholds with numerical solutions
to the full hot-plasma dispersion relation. Section~\ref{sect:density} describes the effects of varying $n_{\alpha}/n_{\mathrm p}$ and the electron temperature $T_{\mathrm e}$. In Section~\ref{sec:contours}, we present analytic fits
to contours of constant growth rate for A/IC and FM/W waves in the $T_{\perp \alpha}/T_{\parallel \alpha}$-$w_{\parallel \alpha}/v_{\mathrm A}$ and $U_{\alpha}/v_{\mathrm A}$-$w_{\parallel \alpha}/v_{\mathrm A}$ planes. In Section~\ref{sect:ql}, we present a
graphical description of the instability mechanism for both wave types
and discuss the way that the alpha particle distribution function
evolves in response to resonant interactions with unstable A/IC waves and FM/W
waves.  We discuss the influence of large-scale magnetic-field-strength fluctuations on the radial evolution of $U_{\alpha}$ in Section~\ref{sect:ani_U} and summarize our conclusions in Section~\ref{sec:conc}.

\section{Resonant Wave--Particle Interactions}\label{sec:framework}

We consider a plasma consisting of protons, electrons, and alpha particles with a background magnetic field of the form $\vec B_0=(0,0,B_0)$. 
 \citet{kennel67} calculated the growth/damping rate of waves with $|\gamma_k|\ll |\omega_{k\mathrm r}|$, where $\omega_{k\mathrm r}$ is the real part of the frequency $\omega_k$ at wavevector $\vec k$ and $\gamma_k$ is its imaginary part. The alpha-particle contribution to $\gamma_k$ is given by
\begin{multline}\label{growthrate}
\frac{\gamma_k^{\alpha}}{|\omega_{k\mathrm r}|}=\frac{\pi}{8n_{\alpha}} \left|\frac{\omega_{k\mathrm r}}{k_{\parallel}} \right|\left(\frac{\omega_{\mathrm p\alpha}}{\omega_{k\mathrm r}}\right)^2 \sum\limits_{n=-\infty}^{+\infty}\int \limits_0^{\infty}\mathrm dv_{\perp} v_{\perp}^2\\ 
\times \int \limits_{-\infty}^{+\infty}\mathrm dv_{\parallel}\,\delta\left(v_{\parallel}-\frac{\omega_{k\mathrm r}-n\Omega_{\alpha}}{k_{\parallel}}\right)\frac{|\psi_{n,k}|^2 \hat Gf_{\alpha}}{W_k},
\end{multline}
where
\begin{equation}\label{waveenerg}
W_k\equiv \frac{1}{16\pi}\left.\left[\vec B^{\ast}_{k}\cdot\vec B_{k}+\vec E^{\ast}_{k}\cdot \diffp{}{\omega}(\omega \varepsilon_{\mathrm h})\vec E_{k}\right]\right|_{\omega=\omega_{k\mathrm r}},
\end{equation}
and $f_{\alpha}$ is the alpha-particle distribution function. The quantities 
\begin{equation}
\vec E_k(\vec k,t)=\int\limits _V \vec E(\vec x,t)e^{-i\vec k\cdot \vec x}\mathrm d^3 x
\end{equation}
and
\begin{equation}
\vec B_k(\vec k,t)=\int\limits_V \vec B(\vec x,t)e^{-i\vec k\cdot \vec x}\mathrm d^3 x
\end{equation}
are the Fourier transforms of the electric and magnetic fields $\vec E(\vec x,t)$ and $\vec B(\vec x,t)$, after these fields have been multiplied by a window function of volume $V$.\footnote{This Fourier transform convention was described in greater detail by \citet{stix92}, although his definitions of $\vec E_k$ and $\vec B_k$ differ from ours by a factor of $\left(2\pi\right)^{-3/2}$.} The left and right circularly polarized components of the electric field are given by $E_{k,\mathrm l}\equiv (E_{kx}+iE_{ky})/\sqrt{2}$ and $E_{k,\mathrm r}\equiv(E_{kx}-iE_{ky})/\sqrt{2}$. 
 $\varepsilon_{\mathrm h}$ denotes the Hermitian part of the dielectric tensor.
 The plasma frequency of the alpha particles is defined by $\omega_{\mathrm p\alpha}^2\equiv 4\pi n_{\alpha}q_{\alpha}^2/m_{\alpha}$, where $q_{\alpha}$ and $m_{\alpha}$ are the alpha-particle charge and mass,
\begin{equation}\label{goperator}
\hat G\equiv\left(1-\frac{k_{\parallel}v_{\parallel}}{\omega_{k\mathrm r}}\right)\diffp{}{v_{\perp}}+\frac{k_{\parallel}v_{\perp}}{\omega_{k\mathrm r}}\diffp{}{v_{\parallel}},
\end{equation}
and
\begin{multline}
\psi_{n,k}\equiv \frac{1}{\sqrt{2}}\left[E_{k,\mathrm r}e^{i\phi}J_{n+1}(x_{\alpha})+E_{k,\mathrm l}e^{-i\phi}J_{n-1}(x_{\alpha})\right]\\
+\frac{v_{\parallel}}{v_{\perp}}E_{kz}J_n(x_{\alpha}).
\label{eq:psink} 
\end{multline}
The argument of the $n$th-order Bessel function $J_n$ is given by $x_{\alpha}=k_{\perp}v_{\perp}/\Omega_{\alpha}$, and $\Omega_{\alpha}=q_{\alpha}B_0/m_{\alpha}c$ is the cyclotron frequency of the alpha particles. 
The velocity is described in cylindrical coordinates with the components $v_{\perp}$  and $v_{\parallel}$ perpendicular and parallel to $\vec B_0$. The wavevector components are given by $k_{\perp}$ and $k_{\parallel}$ in the same geometry, and the azimuthal angle of the wavevector is denoted $\phi$. 
Only waves and particles fulfilling the resonance condition
\begin{equation}\label{rescond}
\omega_{k\mathrm r}=k_{\parallel}v_{\parallel}+n\Omega_{\alpha}
\end{equation}
participate in the resonant wave--particle interaction because of the delta function in Equation~(\ref{growthrate}), which arises formally in \citeauthor{kennel67}'s \citeyearpar{kennel67} analysis because of the condition $|\gamma_k|\ll |\omega_{k\mathrm r}|$.

For concreteness, we take $\omega_{k\mathrm r}>0$ and $k_{\parallel}>0$, preserving full generality by allowing (initially) $U_{\alpha}$ to be either positive, zero, or negative. However, we find below that the plasma is most unstable to drift-anisotropy instabilities when the waves propagate in the same direction as the alpha-particle beam, and so we focus on the case $U_{\alpha}>0$. We assume that $f_{\alpha}$ is  a drifting bi-Maxwellian,
\begin{equation}
f_{\alpha}=\frac{n_{\alpha}}{\pi^{3/2}w_{\perp \alpha}^2w_{\parallel \alpha}} \exp\left(-\frac{v_{\perp}^2}{w_{\perp \alpha}^2}-\frac{\left(v_{\parallel}-U_{\alpha}\right)^2}{w_{\parallel \alpha}^2}\right),
\end{equation}
where
\begin{equation}
w_{\perp \alpha}\equiv \sqrt{\frac{2k_{\mathrm B}T_{\perp \alpha}}{m_{\alpha}}}
\end{equation}
and
\begin{equation}
w_{\parallel \alpha}\equiv \sqrt{\frac{2k_{\mathrm B}T_{\parallel \alpha}}{m_{\alpha}}}
\end{equation}
are the perpendicular and parallel thermal speeds.
The operator $\hat G$ from Equation (\ref{goperator}) applied to the distribution function $f_{\alpha}$ yields
\begin{equation}
\hat Gf_{\alpha}=-\frac{2v_{\perp}f_{\alpha}}{w_{\parallel \alpha}^2 \omega_{k\mathrm r}}\left[\frac{T_{\parallel \alpha}}{T_{\perp \alpha}}\left(\omega_{k\mathrm r}-k_{\parallel}v_{\parallel}\right)+k_{\parallel}\left(v_{\parallel}-U_{\alpha}\right)\right].
\end{equation}
The delta function in Equation (\ref{growthrate}) can be used to evaluate the integral over $v_{\parallel}$ in Equation~(\ref{growthrate}). This delta function can be written in the form $\delta(v_{\parallel}-v_{\mathrm{res}})$, where
\begin{equation}\label{vres}
v_{\mathrm{res}}\equiv \frac{\omega_{k\mathrm r}-n\Omega_{\alpha}}{k_{\parallel}}.
\end{equation}
For the following discussion, we focus on waves with $W_k>0$. The arguments are inverted for negative-energy waves. This effect is, however, not relevant for the parameter range explored in this study \citep[cf.][]{verscharen13}.

\section{Instability Criteria}
\label{sec:criteria}

We develop our analytical model for the case of parallel propagation ($k_{\perp}=0$) only. In this limit, the A/IC wave is left-circularly polarized, and the FM/W wave is right-circularly polarized. For simplicity, we restrict our analysis of the A/IC instability to the case in which $T_{\perp \alpha}\ge T_{\parallel \alpha}$ and our analysis of the FM/W instability to the case in which $T_{\perp \alpha}\le T_{\parallel \alpha}$. Collisions are neglected throughout our treatment.

\subsection{Instability of the Alfv\'en/Ion-Cyclotron (A/IC) Mode}\label{subsect:aic}

In the limit of parallel propagation, the Bessel functions in Equation (\ref{eq:psink}) allow for contributions to the sum over $n$ at $n=1$, $n=-1$, or $n=0$ only. For the left-circularly polarized A/IC wave, $E_{k,\mathrm r}=E_{kz}=0$ and $E_{k,\mathrm l}\neq0$. As a consequence, $\psi_{n,k}=0$ unless $n=1$, and Equation~(\ref{growthrate})  reduces to
\begin{multline}\label{growthlhpol}
\gamma_k^{\alpha}=\frac{1}{16\sqrt{\pi}}\frac{\omega_{k\mathrm r}}{\left|k_{\parallel}w_{\parallel \alpha} \right|} \left(\frac{\omega_{\mathrm p\alpha}}{\omega_{k\mathrm r}}\right)^2\frac{T_{\perp \alpha}}{T_{\parallel \alpha}}\frac{|E_{k,\mathrm l}|^2}{W_k}\\
\times \left[\Omega_{\alpha}\left(1-\frac{T_{\parallel \alpha}}{T_{\perp \alpha}}\right)-\omega_{k\mathrm r}+k_{\parallel}U_{\alpha}\right] \\
\times \exp\left(-\frac{\left(v_{\mathrm{res}}-U_{\alpha}\right)^2}{w_{\parallel \alpha}^2}\right).
\end{multline}
The number of resonant particles is proportional to both $\omega_{\mathrm p\alpha}^2$ and the exponential term in Equation~(\ref{growthlhpol}), and these terms influence the absolute value of $\gamma_k^{\alpha}$. On the other hand, the sign of the growth rate is determined solely by the terms in brackets in Equation (\ref{growthlhpol}), since the other terms are positive semi-definite for $\omega_{k\mathrm r}>0$. We find
\begin{equation}\label{gammakic}
\sgn \gamma_k^{\alpha}=\sgn\left[\Omega_{\alpha}\left(1-\frac{T_{\parallel \alpha}}{T_{\perp \alpha}}\right)-\omega_{k\mathrm r}+k_{\parallel}U_{\alpha}\right].
\end{equation}
This relation defines the maximum frequency at which alpha particles have a destabilizing influence upon the A/IC wave at a given temperature anisotropy, drift, and wavenumber,
\begin{equation}\label{omegamax1}
\omega_{\max}^{\mathrm{A/IC}}=\Omega_{\alpha}\left(1-\frac{T_{\parallel \alpha}}{T_{\perp \alpha}}\right)+k_{\parallel}U_{\alpha}.
\end{equation}
In a plot showing the real-part of the frequency $\omega_{\mathrm r}$ versus $k_{\parallel}$, solutions of the dispersion relation can be driven unstable by alpha particles only when the plot of the dispersion relation is below the line defined by Equation~(\ref{omegamax1}).   The expressions in Equations~(\ref{gammakic}) and (\ref{omegamax1}) are valid for the  general case in which the wave frequency is determined from the full dispersion relation of a hot plasma. However, to simplify the calculation, we now approximate the dispersion relation of the A/IC wave with the dispersion relation of a cold electron-proton plasma with massless electrons,\footnote{We use the equations $k_{\parallel}^2c^2=\omega^2L$ for the A/IC wave and  $k_{\parallel}^2c^2=\omega^2R$ for the FM/W wave following the notation of \citet{stix92}.  Equation~(\ref{genICdisp}) follows,
for example, from Equation (5) in Chapter 2 of \citet{stix92} under the
approximations that $\omega \ll \Omega_{\mathrm e}$ and $v_{\mathrm A} \ll
c$. Details on modifications of the dispersion relations of the two parallel normal modes due to alpha particles and finite-pressure effects are described in the literature \citep[see e.g.,][]{marsch11,verscharen11,thesis}.}
\begin{equation}\label{genICdisp}
\frac{\omega_{k\mathrm r}}{\Omega_{\mathrm p}}=-\frac{k_{\parallel}^2v_{\mathrm A}^2}{2\Omega_{\mathrm p}^2}\left[1-\sqrt{1+\frac{4\Omega_{\mathrm p}^2}{k_{\parallel}^2v_{\mathrm A}^2}}\right].
\end{equation}
When $k_{\parallel}v_{\mathrm A}\ll \Omega_{\mathrm p}$, Equation~(\ref{genICdisp}) can be approximated as
\begin{equation}\label{approxDRLH}
\omega_{k\mathrm r}\simeq k_{\parallel}v_{\mathrm A}\left(1-\frac{k_{\parallel}v_{\mathrm A}}{2\Omega_{\mathrm p}}\right).
\end{equation}
As we will describe further below, the wavenumbers at which the A/IC mode is unstable for typical solar-wind parameters near 1 AU are $\lesssim 0.5\Omega_{\mathrm p}/v_{\mathrm A}$. To further simplify our analysis, we will thus use Equation~(\ref{approxDRLH}) instead of Equation~(\ref{genICdisp}).

\begin{figure}
\epsscale{1.2}
\plotone{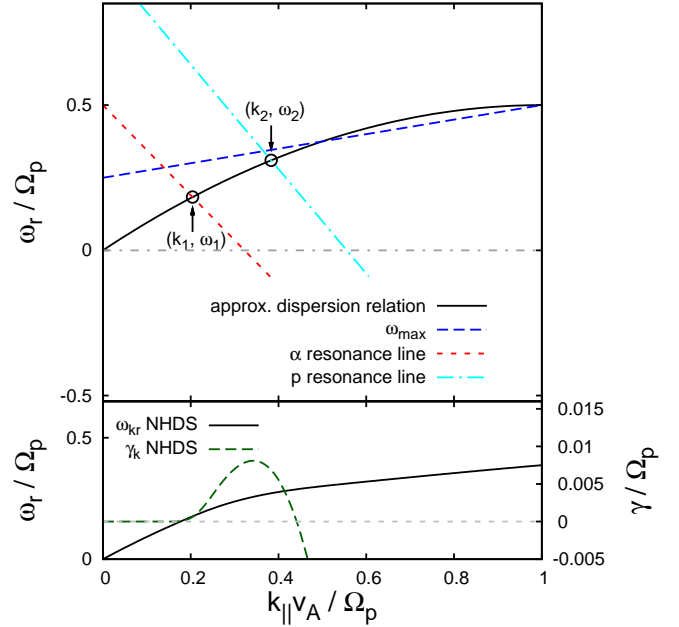}
\caption{Resonances and instability conditions for the Alfv\'en/ion-cyclotron (A/IC) mode. The upper diagram shows the approximate dispersion relation, resonance lines, and the quantities $k_1$, $k_2$, $\omega_1$, and $\omega_2$ discussed in the text for the case in which $U_{\alpha}=0.25v_{\mathrm A}$, $T_{\perp \alpha }/T_{\parallel \alpha }=2$, $w_{\parallel \alpha}=0.75v_{\mathrm A}$, and $\sigma=2.4$. The lower panel shows solutions of the full dispersion relation of a hot plasma for the same parameters with the real part of the frequency on the left axis and the corresponding growth rate on the right axis. The additional parameters are $n_{\alpha}=0.05 n_{\mathrm p}$ and $T_{\mathrm p}=T_{\mathrm e}=T_{\parallel \alpha}/4$. Protons and electrons have no temperature anisotropy.}
\label{fig_ic-dati}
\end{figure}

At the resonant velocity $v_{\parallel}=v_{\mathrm{res}}$, a large enough number of particles have to be present in the distribution function to drive the wave unstable.
This requires that
\begin{equation}\label{defdeltav}
\left|v_{\parallel}-U_{\alpha}\right|<\sigma w_{\parallel \alpha}\equiv \Delta v_{\parallel},
\end{equation}
where $\sigma$ is a number that can not be much greater than unity. 
We define the \emph{resonance line} of the alpha particles through the equation
\begin{equation}\label{alpharesline}
\omega_{\mathrm r}=k_{\parallel}(U_{\alpha}-\Delta v_{\parallel})+\Omega_{\alpha}=k_{\parallel}(U_{\alpha}-\sigma w_{\parallel \alpha})+\Omega_{\alpha}.
\end{equation}
It represents the resonance condition Equation~(\ref{rescond}) for alpha particles at the parallel velocity $v_{\parallel}=U_{\alpha}-\Delta v_{\parallel}$. 
The intersection of the alpha-particle resonance line in the $\omega_{\mathrm r}$-$k_{\parallel}$ plane with the plot of the dispersion relation of the A/IC wave defines the point $(k_1, \omega_1)$ as illustrated in Figure~\ref{fig_ic-dati}. 
Only solutions of the dispersion relation at $k_{\parallel}>k_1$ and $\omega_{k\mathrm r}>\omega_1$ are able to interact resonantly with a sufficiently large number of alpha particles to excite an instability. 

Although we have only quoted the expression for the alpha-particle contribution to $\gamma_k$ in Section~\ref{sec:framework}, the proton contribution is given by an analogous expression \citep{kennel67}, in which the resonance condition becomes $\omega_{k\mathrm r}=k_{\parallel}v_{\parallel}+n\Omega_{\mathrm p}$, where $\Omega_{\mathrm p}=2\Omega_{\alpha}$ is the proton cyclotron frequency. We assume that the protons have a Maxwellian distribution, which means that they damp A/IC waves if they can satisfy the resonance condition with $n=1$. Because the protons are the dominant ion species, we assume that if thermal protons can resonate with A/IC waves, then proton damping dominates over any possible wave destabilization by the alpha particles.
 We define the resonance line of the protons through the equation
\begin{equation}\label{protres}
\omega_{\mathrm r}=- k_{\parallel} \sigma w_{\parallel \mathrm p}+\Omega_{\mathrm p},
\end{equation}
where $w_{\parallel \mathrm p}$ is the parallel proton thermal speed. 
Equation~(\ref{protres}) is the $n=1$ resonance condition for protons with $v_{\parallel}=-\sigma w_{\parallel \mathrm p}$. For simplicity, in this section we take $\sigma$ to be the same for alpha particles and protons for our fiducial case in which $n_{\alpha}=0.05n_{\mathrm p}$. However, in Section~\ref{sect:density} we describe how $\sigma$ should be varied for alpha particles as $n_{\alpha}/n_{\mathrm p}$ is varied.
We plot the proton resonance line in Figure~\ref{fig_ic-dati} for the case in which $w_{\parallel \mathrm p}=w_{\parallel \alpha}$, which is characteristic of weakly collisional solar-wind streams at 1 AU \citep{kasper08}. The intersection of this resonance line and the plot of the dispersion relation in the $\omega_{\mathrm r}$-$k_{\parallel}$ plane defines an upper limit for the unstable wavenumbers and frequencies. We denote this intersection point as $(k_2, \omega_2)$ as illustrated in Figure~\ref{fig_ic-dati}.

The above considerations lead to the following necessary and sufficient conditions for the A/IC wave to be unstable:
\begin{enumerate}
\item There must exist a range of wavenumbers in which thermal alpha particles can resonate with the A/IC wave that is sufficiently small that proton cyclotron damping can be neglected: i.e., $k_1$ exists, and $k_1 < k_2$. 
\item In at least part of the wavenumber interval $[k_1, k_2]$, the wave frequency $\omega_{k\mathrm r}$  must be less than the maximum wave frequency given by $\omega_{\max}^{\mathrm{A/IC}}$.
\end{enumerate}
We henceforth restrict our analysis to the case in which
\begin{equation}\label{k3defapp}
\sigma w_{\parallel \mathrm p} > \frac{v_{\mathrm A}}{2}.
\end{equation}
Equation~(\ref{k3defapp}) implies that $k_2 v_{\mathrm A}/\Omega_{\mathrm p} < 1$, so that proton
cyclotron damping limits potential instabilities to parallel wavenumbers less than $ \Omega_{\mathrm p}/v_{\mathrm A}$, for which our approximate dispersion
relation is at least approximately valid. 

After some algebra, we find
that condition 1 above is equivalent to the inequality
\begin{equation}\label{cond3b}
\sigma w_{\parallel \alpha}>\frac{4U_{\alpha}+3\sigma w_{\parallel \mathrm p}-v_{\mathrm A} -\sqrt{\left(\sigma w_{\parallel \mathrm p}+v_{\mathrm A}\right)^2-2v_{\mathrm A}^2}}{4}.
\end{equation}
Condition 2 is satisfied if either
\begin{equation}\label{cond1b}
U_{\alpha}>v_{\mathrm A}-\sigma \left(\frac{T_{\perp \alpha}}{T_{\parallel \alpha}}-1\right)w_{\parallel \alpha}-\frac{v_{\mathrm A}^2T_{\parallel \alpha}}{4\sigma w_{\parallel \alpha}T_{\perp \alpha}}
\end{equation}
and
\begin{equation}\label{cond2b}
\sigma w_{\parallel \alpha}>\frac{v_{\mathrm A}T_{\parallel \alpha}}{2T_{\perp \alpha}},
\end{equation}
or if
\begin{equation}\label{eq60}
U_{\alpha}>\frac{T_{\perp \alpha}+T_{\parallel \alpha}}{4T_{\perp \alpha}}\left[\sigma w_{\parallel \mathrm p}+v_{\mathrm A}+\sqrt{\left(v_{\mathrm A}+\sigma w_{\parallel \mathrm p}\right)^2-2v_{\mathrm A}^2}\right]-\sigma w_{\parallel \mathrm p}.
\end{equation}
Because the wave frequency $\omega_{k\mathrm r}$ in Equation~(\ref{approxDRLH}) is concave downwards ($d^2 \omega_{k\mathrm r} /dk_{\parallel}^2 < 0 $), the condition that $\omega_{k\mathrm r} < \omega_{\max}^{\mathrm{A/IC}}$ within the interval $[k_1, k_2]$ is most easily satisfied at either $k_{\parallel}=k_1$ or at $k_{\parallel}=k_2$. Equations~(\ref{cond1b}) and (\ref{cond2b}) are the conditions that $\omega_{k\mathrm r} < \omega_{\max}^{\mathrm{A/IC}}$ at $k_{\parallel}=k_1$, while Equation~(\ref{eq60}) is the condition that $\omega_{k\mathrm r} < \omega_{\max}^{\mathrm{A/IC}}$ at $k_{\parallel}=k_2$. In the Appendix, we show which of these two alternative conditions is less restrictive (and therefore the controlling instability criterion)
as a function of $w_{\parallel \alpha}/v_{\mathrm A}$, $w_{\parallel \mathrm p}/v_{\mathrm A}$, and $T_{\perp \alpha}/T_{\parallel \alpha}$.
Roughly speaking, for typical solar wind conditions near $r= 1 \mbox{ AU}$, Equations~(\ref{cond1b}) and (\ref{cond2b}) are less restrictive when $T_{\perp \alpha}\gtrsim 1.1 T_{\parallel \alpha}$, while Equation~(\ref{eq60}) is less restrictive for nearly isotropic alpha-particle temperatures. To apply these conditions, one can either consult the conditions given in the Appendix, or 
take the minimum $U_\alpha$ needed to excite the instability to be the minimum of the right-hand sides of Equations~(\ref{cond1b}) and (\ref{eq60}) (with the caveat that Equation~(\ref{cond1b}) applies only when Equation~(\ref{cond2b}) is also satisfied).

For reference, we note that Equation~(\ref{cond1b}) can be rewritten as a lower limit on the temperature anisotropy,
\begin{equation}\label{cond1ani}
\frac{T_{\perp \alpha}}{T_{\parallel \alpha}}>\frac{v_{\mathrm A}+\sigma w_{\parallel \alpha}-U_{\alpha}+\sqrt{\left(v_{\mathrm A}+\sigma w_{\parallel \alpha}-U_{\alpha}\right)^2-v_{\mathrm A}^2}}{2\sigma w_{\parallel \alpha}}.
\end{equation} 
Equation~(\ref{eq60}) can be equivalently rewritten as an upper limit on the inverse temperature anisotropy,
\begin{multline}\label{eq60ani}
\frac{T_{\parallel \alpha}}{T_{\perp \alpha}}<\frac{2}{v_{\mathrm A}^2}\left(U_{\alpha}+\sigma w_{\parallel \mathrm p}\right)\\
\times \left[v_{\mathrm A}+\sigma w_{\parallel \mathrm p}-\sqrt{\left(v_{\mathrm A}+\sigma w_{\parallel \mathrm p}\right)^2-2v_{\mathrm A}^2}\right]-1.
\end{multline}
Likewise, Equation~(\ref{cond3b}) can be rewritten as an upper limit on the drift speed,
\begin{equation}\label{upperlimitDCI}
U_{\alpha}<\frac{4\sigma w_{\parallel \alpha}-3\sigma w_{\parallel \mathrm p}+v_{\mathrm A}+\sqrt{\left(v_{\mathrm A}+\sigma w_{\parallel \mathrm p}\right)^2-2v_{\mathrm A}^2}}{4}.
\end{equation}

In addition to the illustrations for our analytical model, Figure~\ref{fig_ic-dati} shows solutions of the full dispersion relation of a hot plasma for the A/IC mode in the lower panel.  These results have been obtained with the NHDS code \citep{verscharen13a}. The parameters for the NHDS calculation are $T_{\parallel \alpha}=4T_{\mathrm p}=4T_{\mathrm e}$, $n_{\alpha}=0.05n_{\mathrm p}$, $T_{\perp \alpha}=2T_{\parallel \alpha}$, $T_{\perp}=T_{\parallel}$ for electrons and protons, $U_{\alpha}=0.25v_{\mathrm A}$, and $w_{\parallel \alpha}=0.75v_{\mathrm A}$.
In all numerical calculations presented throughout the text, we set
\begin{equation}
 v_{\mathrm A}/c=10^{-4}
\end{equation}
and choose the electron number density and electron bulk drift velocity so that the net space charge and net parallel current vanish. The onset of the instability (i.e., the wavenumber range where $\gamma_k>0$) coincides well with the predictions from our analytical model. 

\begin{figure}
\epsscale{1.2}
\plotone{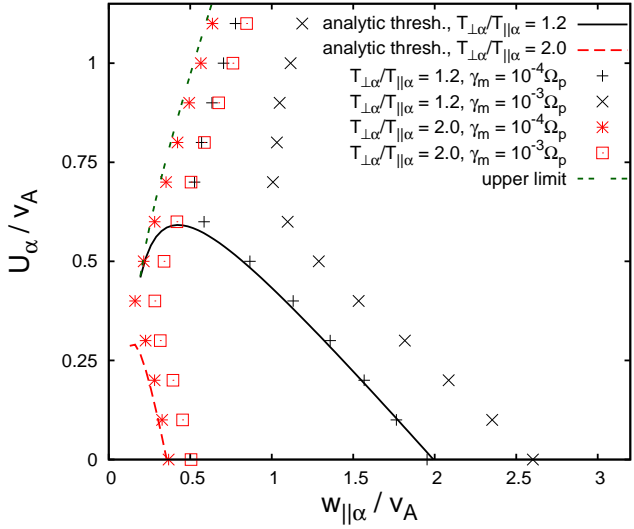}
\caption{Comparison of the instability criteria and numerical solutions to the hot-plasma dispersion relation for the A/IC mode. The black  and red dashed lines correspond to Equation~(\ref{cond1b}) for two different temperature anisotropies. The green double dashed line shows the upper limit on $U_{\alpha}$ from Equation~(\ref{upperlimitDCI}). We set $\sigma=2.4$. The points represent solutions of the hot-plasma dispersion relation from the NHDS code with the given maximum growth rate $\gamma_{\mathrm m}$. The density of the alpha particles is $n_{\alpha}=0.05n_{\mathrm p}$, $T_{\mathrm p}=T_{\mathrm e}=T_{\parallel \alpha}/4$, and protons and electrons have isotropic distribution functions in the NHDS solutions.}
\label{fig_ic-dati-Uw}
\end{figure}

We compare the analytical instability conditions from Equations~(\ref{cond1b}) and (\ref{cond2b}) with solutions of the full dispersion relation of a hot plasma in Figure~\ref{fig_ic-dati-Uw}. We have adjusted the single free parameter $\sigma$ to maximize the agreement between our analytic and numerical results, and find that $\sigma=2.4$ leads to the best fit with the numerical solutions corresponding to $\gamma_{\mathrm m}=10^{-4} \Omega_{\mathrm p}$ (but see Section~\ref{sect:density}). 
We note that the wavenumbers at which the maximum growth rates occur in the NHDS solutions are less than $0.7\Omega_{\mathrm p}/v_{\mathrm A}$.

\begin{figure}
\epsscale{1.2}
\plotone{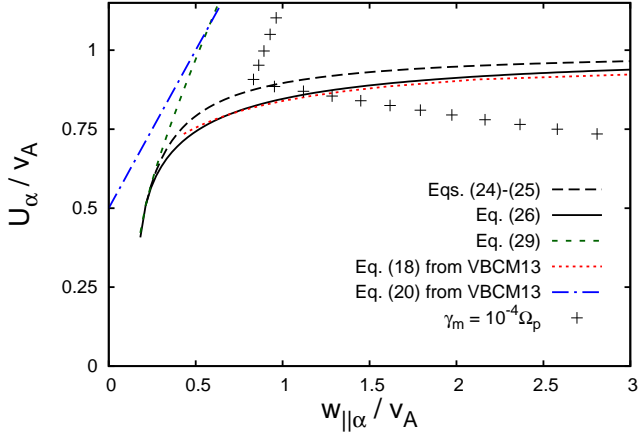}
\caption{Comparison of the instability criteria and numerical solutions to the hot-plasma dispersion relation for the A/IC mode in the isotropic case. The label ``VBCM13'' refers to the treatment of the isotropic parallel Alfv\'enic instability by \citet{verscharen13a}.  Equation~(\ref{eq60}) leads to a lower $U_{\alpha}$ threshold than Equation~(\ref{cond1b}). The points represent solutions of constant $\gamma_{\mathrm m}$ obtained from the NHDS code. We set $\sigma=2.4$. For the NHDS solutions,  $n_{\alpha}=0.05n_{\mathrm p}$, $T_{\mathrm p}=T_{\mathrm e}=T_{\parallel \alpha}/4$, and protons and electrons have Maxwellian distribution functions.}
\label{fig_ic-dati-Uweq60}
\end{figure}

Equation~(\ref{eq60}) describes a lower threshold than Equations~(\ref{cond1b}) and (\ref{cond2b}) in the isotropic limit. We show this case in Figure~\ref{fig_ic-dati-Uweq60}, in which we also compare the threshold with the previous isotropic-temperature model developed by \citet{verscharen13a} and with NHDS solutions with $\gamma_{\mathrm m}=10^{-4}\Omega_{\mathrm p}$.
We see that the lower and upper limits described by our instability criteria agree well with the previous model, which is based on two alternative approximations to the dispersion relation. We note that the difference between the instability thresholds from Equations~(\ref{cond1b}) and (\ref{eq60}) is small. The difference between both analytic expressions is in most cases smaller than the difference between either of them and the solution to the hot-plasma dispersion relation. We find a similar behavior in all tested cases for which Equation~(\ref{eq60}) describes a lower threshold. At high parallel thermal speeds $w_{\parallel \alpha}$, the inaccuracies of our approximate dispersion relation increase.
For a more extensive discussion of Equation~(\ref{eq60}) and the comparison between the two thresholds, we refer the reader again to the Appendix.

\subsection{Instability of the Fast-magnetosonic/Whistler (FM/W) Mode}\label{sect:DFI}

For the right-circularly polarized FM/W mode, $E_{k,\mathrm l}=E_{kz}=0$ and $E_{k,\mathrm r}\neq 0$. As a result, $\psi_{n,k}=0$ unless $n=-1$, and Equation~(\ref{growthrate}) reduces to 
\begin{multline}\label{growthrhpol}
\gamma_k^{\alpha}=\frac{1}{16\sqrt{\pi}}\frac{\omega_{k\mathrm r}}{\left|k_{\parallel}w_{\parallel \alpha} \right|} \left(\frac{\omega_{\mathrm p\alpha}}{\omega_{k\mathrm r}}\right)^2\frac{T_{\perp \alpha}}{T_{\parallel \alpha}}\frac{|E_{k,\mathrm r}|^2}{W_k}\\
\times \left[\Omega_{\alpha}\left(\frac{T_{\parallel \alpha}}{T_{\perp \alpha}}-1\right)-\omega_{k\mathrm r}+k_{\parallel}U_{\alpha}\right] \\
\times \exp\left(-\frac{\left(v_{\mathrm{res}}-U_{\alpha}\right)^2}{w_{\parallel \alpha}^2}\right).
\end{multline}
It follows that
\begin{equation}\label{gammasignFH}
\sgn \gamma_k^{\alpha}=\sgn\left[\Omega_{\alpha}\left(\frac{T_{\parallel \alpha}}{T_{\perp \alpha}}-1\right)-\omega_{k\mathrm r}+k_{\parallel}U_{\alpha}\right].
\end{equation}
Equation~(\ref{gammasignFH}) defines the maximum frequency at which alpha particles have a destabilizing influence on the FM/W wave at a given temperature anisotropy, drift, and wavenumber,
\begin{equation}\label{omegamax2}
\omega_{\max}^{\mathrm{FM/W}}=\Omega_{\alpha}\left(\frac{T_{\parallel \alpha}}{T_{\perp \alpha}}-1\right)+k_{\parallel}U_{\alpha}.
\end{equation}
Equation~(\ref{rescond}) implies that the resonant alpha particles that drive the FM/W mode unstable have a parallel velocity $v_{\parallel}>0$, since we have assumed that $\omega_{k\mathrm r}>0$. 

To simplify the analysis, we approximate $\omega_{k\mathrm r}$ using the dispersion relation for right-circularly polarized FM/W waves in a cold proton-electron plasma with massless electrons,\footnote{Equation (\ref{fulldispDFI}) follows, for example, from Equation (5) in Chapter 2 of \citet{stix92} under the approximations that $\omega \ll \Omega_{\mathrm e}$ and $v_{\mathrm A} \ll c$.}
\begin{equation}\label{fulldispDFI}
\frac{\omega_{k\mathrm r}}{\Omega_{\mathrm p}}=\frac{k_{\parallel}^2v_{\mathrm A}^2}{2\Omega_{\mathrm p}^2}\left[1+\sqrt{1+\frac{4\Omega_{\mathrm p}^2}{k_{\parallel}^2v_{\mathrm A}^2}}\right].
\end{equation}
We further simplify $\omega_{k\mathrm r}$ by taking $k_{\parallel}v_{\mathrm A}/\Omega_{\mathrm p}\ll 1$ and expanding Equation~(\ref{fulldispDFI}) as
\begin{equation}\label{appdispRH}
\omega_{k\mathrm r}\simeq k_{\parallel}v_{\mathrm A}\left(1+\frac{k_{\parallel}v_{\mathrm A}}{2\Omega_{\mathrm p}}\right).
\end{equation}

\begin{figure}
\epsscale{1.2}
\plotone{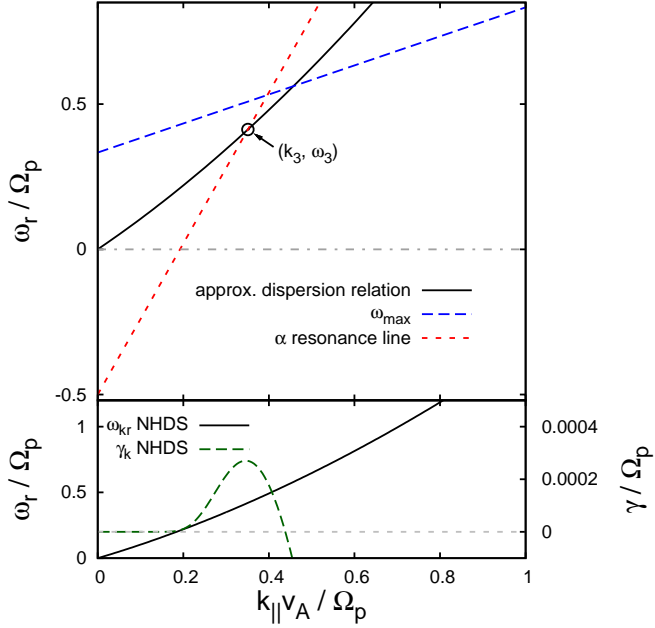}
\caption{Upper diagram: resonances and instability conditions for the fast-magnetosonic/whistler (FM/W) mode for the case in which $U_{\alpha}=0.5v_{\mathrm A}$, $T_{\perp \alpha }/T_{\parallel \alpha }=0.6$, $w_{\parallel \alpha}=v_{\mathrm A}$, and $\sigma=2.1$. Lower diagram: numerical solutions of the full dispersion relation of a hot plasma for the same parameters with the real part of the frequency on the left axis and the corresponding growth rate on the right axis. The additional parameters are $n_{\alpha}=0.05 n_{\mathrm p}$, $T_{\mathrm p}=T_{\mathrm e}=T_{\parallel \alpha}/4$. Protons and electrons have no temperature anisotropy.}
\label{fig_fh-dati}
\end{figure}

 From the resonance condition Equation~(\ref{rescond}), we conclude that there is a minimum value of $v_{\parallel}$ for which alpha particles can resonate with the FM/W wave. Again there must be a sufficient number of alpha particles at that $v_{\parallel}$ to excite the instability. We make the approximation that the instability is possible only if alpha particles with $v_{\parallel}=U_{\alpha}+\sigma w_{\parallel \alpha}$ resonate with the wave, where $\sigma$ is a constant of order unity, as in Section~\ref{subsect:aic}. We therefore define the alpha-particle resonance line for the FM/W wave through the equation 
\begin{equation}
\omega_{\mathrm r}=k_{\parallel}(U_{\alpha}+\sigma w_{\parallel \alpha})-\Omega_{\alpha}.
\end{equation}
Its intersection with the plot of the dispersion relation is denoted $(k_3, \omega_3)$, as illustrated in Figure~\ref{fig_fh-dati}.

As in the case of the A/IC wave, the necessary and sufficient conditions for instability of the FM/W wave consist of two criteria. First, there must be some range of wavenumbers exceeding $k_3$ in which the wave frequency $\omega_{k\mathrm r}$ is less than the maximum wave frequency $\omega_{\max}^{\mathrm{FM/W}}$. We write this condition after some algebra as
\begin{equation}\label{lowerlimitDFI}
U_{\alpha}>v_{\mathrm A}-\sigma \left(1-\frac{T_{\perp \alpha}}{T_{\parallel \alpha}}\right)w_{\parallel \alpha}+\frac{v_{\mathrm A}^2T_{\parallel \alpha}}{4\sigma w_{\parallel \alpha}T_{\perp \alpha}}.
\end{equation}
We can rewrite Equation~(\ref{lowerlimitDFI}) as a minimum threshold for the temperature anisotropy at a given drift speed:
\begin{equation}\label{cond2}
\frac{T_{\perp \alpha}}{T_{\parallel \alpha}}<\frac{U_{\alpha}+\sigma w_{\parallel \alpha}-v_{\mathrm A}+\sqrt{\left(U_{\alpha}+\sigma w_{\parallel \alpha}-v_{\mathrm A}\right)^2-v_{\mathrm A}^2}}{2\sigma w_{\parallel \alpha}}.
\end{equation}
The point $(k_3,\omega_3)$ is only defined if 
\begin{equation}\label{k2defFH}
U_{\alpha}>2v_{\mathrm A}-\sigma w_{\parallel \alpha}.
\end{equation}
If Equation~(\ref{k2defFH}) is not fulfilled, the alpha-particle resonance line does not intersect with the plot of the dispersion relation in the $\omega_{\mathrm r}$-$k_{\parallel}$ plane, and the square-root expression in Equation~(\ref{cond2}) is not real.

The second condition is that the proton thermal speed must be sufficiently small that proton damping can be neglected within some part of the wavenumber interval defined in the previous paragraph (in which $k_{\parallel}>k_3$ and $\omega_{k\mathrm r}<\omega_{\max}^{\mathrm{FM/W}}$). Because the FM/W wave is right-circularly polarized, proton damping occurs through the $n=-1$ resonance. As in our treatment of the A/IC wave, we assume that if thermal protons can resonate with the FM/W wave, then proton damping dominates over any possible destabilizing influence from the resonant alpha particles. This second condition can be written in the form
\begin{equation}\label{protdamp}
\sigma w_{\parallel \mathrm p} <\frac{  3U_{\alpha}+3\sigma w_{\parallel \alpha}-v_{\mathrm A} + \sqrt{\left(U_{\alpha}+\sigma w_{\parallel \alpha}-v_{\mathrm A}\right)^2 - v_{\mathrm A}^2}}{2}.
\end{equation}
In this paper, we focus on the case in which $w_{\parallel \mathrm p}=w_{\parallel \alpha}$. The condition  $w_{\parallel \mathrm p}\le w_{\parallel \alpha}$  and $U_{\alpha}\ge 0$ guarantee that Equation~(\ref{protdamp}) is satisfied. 

We show NHDS solutions for the FM/W mode in the lower panel of Figure~\ref{fig_fh-dati}. The range of unstable wavenumbers roughly corresponds to the predictions from our analytical model. At wavenumbers exceeding $\sim 0.45\Omega_{\mathrm p}/v_{\mathrm A}$, $\omega_{k\mathrm r}>\omega_{\max}^{\mathrm{FM/W}}$ and, therefore, the interaction with the alpha particles leads to a damping instead of an instability. In contrast to the A/IC wave, the instability of the FM/W wave involves only a lower limit on $U_{\alpha}$.

\begin{figure}
\epsscale{1.2}
\plotone{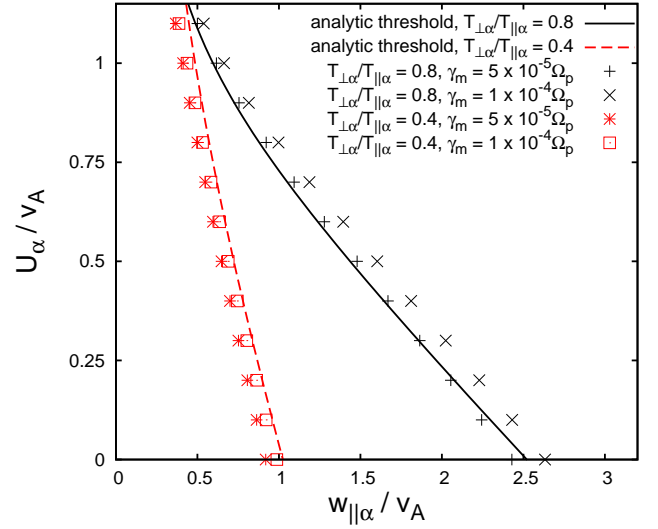}
\caption{Comparison of the analytic instability criteria and numerical solutions to the hot-plasma dispersion relation for the FM/W mode. The black solid and red dashed lines show the analytical result for the stability thresholds according to Equation~(\ref{lowerlimitDFI}) for two different temperature anisotropies. We set $\sigma=2.1$. The points represent solutions of the hot-plasma dispersion relation obtained by NHDS at the given maximum growth rate $\gamma_{\mathrm m}$. The density of the alpha particles is $n_{\alpha}=0.05n_{\mathrm p}$, $T_{\mathrm p}=T_{\mathrm e}=T_{\parallel \alpha}/4$, and protons and electrons have Maxwellian distribution functions for the NHDS evaluation.}
\label{fig_fh-dati-Uw}
\end{figure}

In Figure~\ref{fig_fh-dati-Uw}, we plot Equation~(\ref{lowerlimitDFI}) for two different values of $T_{\perp \alpha}/T_{\parallel \alpha}$, along with the locations of fixed maximum growth rates in numerical solutions to the full hot-plasma dispersion relation obtained with the NHDS code. As in Section~\ref{subsect:aic}, we vary the parameter $\sigma$, and find that the value $\sigma=2.1$ leads to the best fit between the analytic threshold and NHDS solutions with $\gamma_{\mathrm m}=10^{-4}\Omega_{\mathrm p}$ for the FM/W instability, especially in the range of low $w_{\parallel \alpha}/v_{\mathrm A}$ where our approximate dispersion relation is more accurate. 
 The NHDS solutions for the maximum growth rate of the instability of the FM/W mode occur at wavenumbers below $k_{\parallel}v_{\mathrm A}/\Omega_{\mathrm p}\sim 0.8$. This justifies our expansion of Equation~(\ref{fulldispDFI}) to obtain Equation~(\ref{appdispRH}).

\begin{figure}
\epsscale{1.2}
\plotone{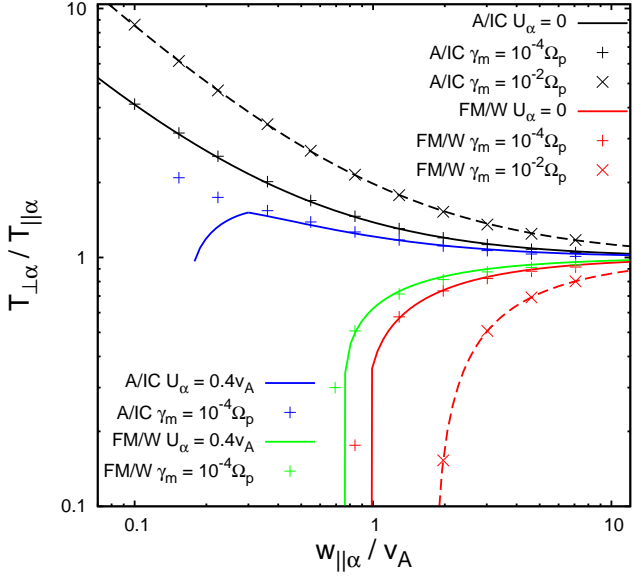}
\caption{Instability thresholds in the $T_{\perp \alpha}/T_{\parallel \alpha}$-$w_{\parallel \alpha}/v_{\mathrm A}$ plane for $U_{\alpha}=0$ (black and red lines) and $U_{\alpha}=0.4v_{\mathrm A}$ (blue and green lines). We show the lowest analytical thresholds from Equations~(\ref{cond1ani}) and (\ref{eq60ani}) for the A/IC mode, and the analytical threshold from (\ref{cond2}) for the FM/W mode as solid lines. We set $\sigma=2.4$ for the A/IC instability and $\sigma=2.1$ for the FM/W instability. The dashed lines show fit results from \citet{maruca12} as isocontours for a maximum growth rate $\gamma_{\mathrm m}=10^{-2}\Omega_{\mathrm p}$. The points show NHDS results for different growth rates. The parameters for the NHDS calculation are $n_{\alpha}=0.05n_{\mathrm p}$, $T_{\mathrm p}=T_{\mathrm e}=T_{\parallel \alpha}/4$. Protons and electrons have Maxwellian distribution functions.}
\label{fig_dati_baleplot}
\end{figure}

Another way of illustrating the instability thresholds is shown in Figure~\ref{fig_dati_baleplot}. In this figure, we compare our analytic thresholds for the A/IC wave and the FM/W wave (Equations~(\ref{cond1ani}), (\ref{eq60ani}), and (\ref{cond2})) with NHDS solutions to the full hot-plasma dispersion relation for the cases $U_{\alpha}=0$ and $U_{\alpha}=0.4v_{\mathrm A}$.
We find a very good agreement between our analytical thresholds and the NHDS solutions at low growth rates. The presence of a non-zero drift $U_{\alpha}$ lowers the thresholds in terms of $|T_{\perp \alpha}/T_{\parallel \alpha}-1|$ and narrows the stable parameter range.  By applying the NHDS code to find parameter combinations that lead to $\gamma_{\mathrm m}=10^{-2}\Omega_{\mathrm p}$,  we are able to confirm the previous fit results by \citet{maruca12} for $T_{\parallel \alpha}=4T_{\parallel \mathrm p}$ and $T_{\perp \mathrm p}=T_{\parallel \mathrm p}$, which are also included in Figure~\ref{fig_dati_baleplot}. The comparison also shows in which parameter ranges the isocontours of constant maximum growth rate have larger distances and where they have shorter distances from each other. The FM/W instability shows a very sensitive dependence to the assumed growth rate. Our threshold for the instability of the FM/W mode drops abruptly at the point where Equation~(\ref{k2defFH}) is violated. We expect a smoother transition in a real plasma since the number of resonant particles does not drop to zero at a discrete value of $v_{\parallel}$. The location of the point where the analytical threshold drops depends on the assumed value of $\sigma$.

\section{The Dependence on the Alpha-Particle Number Density and the Electron Temperature}\label{sect:density}

\begin{figure}
\epsscale{1.2}
\plotone{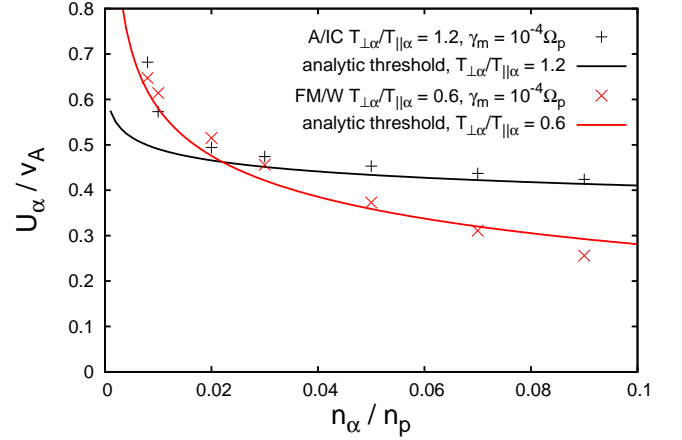}
\caption{Dependence of the instability thresholds on $n_{\alpha}/n_{\mathrm p}$ in the case in which $w_{\parallel \mathrm p}=w_{\parallel \alpha}=v_{\mathrm A}$ and $T_{\mathrm p}=T_{\mathrm e}$. The points represent NHDS solutions at two different growth rates of the A/IC and the FM/W modes. The curves correspond to the analytical thresholds from Equations~(\ref{cond1b}) and (\ref{lowerlimitDFI}) with $\sigma$ of the alphas described by Equation~(\ref{sigmaevolution}). }
\label{fig_density_dependence_gamma}
\end{figure}

In the previous sections, we assumed that $\sigma$ does not depend upon plasma parameters and determined the values of $\sigma$ by comparing our analytic thresholds to numerical solutions of the hot-plasma dispersion relation for a single value of $n_{\alpha}/n_{\mathrm p}$. We conjecture, however, that as $n_{\alpha}/n_{\mathrm p}$ varies, the number of resonant particles that are needed in order to achieve an instability (as a fraction of $n_{\mathrm p}$) remains constant, i.e., $(n_{\alpha}/n_{\mathrm p})\exp(-\sigma^2)\equiv M=\mathrm{constant}$. A good value for this constant determined from the comparison of our analytical thresholds and the hot-plasma dispersion relation at $n_{\alpha}=0.05n_{\mathrm p}$ is given by 
\begin{equation}
M=\left\{\begin{array}{ll}1.6\times 10^{-4} & \,\,\text{for the A/IC wave}\\ 6.1\times 10^{-4} & \,\,\text{for the FM/W wave}\end{array}\right.
\end{equation}
This requirement leads to a density-dependence of the factor $\sigma$ for the alpha particles of the form
\begin{equation}\label{sigmaevolution}
\sigma=\sqrt{-\ln \frac{Mn_{\mathrm p}}{n_{\alpha}}}.
\end{equation} 
For the protons, we continue to set $\sigma=2.4$ for the A/IC wave and $\sigma=2.1$ for the FM/W wave as $n_{\alpha}/n_{\mathrm p}$ varies. Thus, Equation~(\ref{sigmaevolution}) is used for each factor of $\sigma$ that multiplies the quantity $w_{\parallel \alpha}$ in any of our equations, but the values $\sigma=2.4$ and $\sigma=2.1$, respectively, are used for each value of $\sigma$ that multiplies the quantity $w_{\parallel \mathrm p}$.
We illustrate the dependence of the thresholds on the fractional alpha-particle density for two sets of parameters in Figure~\ref{fig_density_dependence_gamma}.
We show NHDS solutions for a maximum growth rate of $\gamma_{\mathrm m}= 10^{-4}\Omega_{\mathrm p}$ for the A/IC instability and the FM/W instability. We compare the NHDS points with the analytical instability thresholds from Equations~(\ref{cond1b}) and (\ref{lowerlimitDFI}). The thresholds plotted in Figure~\ref{fig_density_dependence_gamma} have been calculated using Equation~(\ref{sigmaevolution}). 

In the fast solar wind, $n_{\alpha}/n_{\mathrm p}\simeq 0.05$ \citep{bame77}, while in the slow solar wind $n_{\alpha}/n_{\mathrm p}$ is roughly proportional to the wind speed, with $0.01\lesssim n_{\alpha}/n_{\mathrm p}\lesssim 0.05$ \citep{kasper07}. Within the range $0.01<n_{\alpha}/n_{\mathrm p}<0.05$, the instability thresholds shown in Figure~\ref{fig_density_dependence_gamma} exhibit a moderate density dependence, which is captured fairly accurately by taking $\sigma$ fixed at the values 2.4 (A/IC) or 2.1 (FM/W) for the protons and using Equation~(\ref{sigmaevolution}) for the alpha particles. We conjecture that this approach for modeling $\sigma$ approximately captures the density dependence of the instability thresholds more generally, at least when $n_{\alpha}/n_{\mathrm p}\in (0.01,0.05)$. When $n_{\alpha}/n_{\mathrm p} >0.05$, the alpha particles increasingly modify the dispersion relation, and we expect our analytic results to become less accurate.

Another parameter that could in principle influence the thresholds of the instability is the ratio $T_{\mathrm e}/T_{\mathrm p}$. However, we expect that this temperature ratio has a negligible influence on both instabilities since the corresponding waves have $E_{kz}=0$ and $k_{\perp}=0$, which means that electrons cannot interact with these waves through Landau or transit-time damping. Cyclotron-resonant interaction is also not expected to influence the wavenumber range where alpha-particle driving is important because $\Omega_{\mathrm e}\gg \Omega_{\alpha}$. 
NHDS calculations at different $T_{\mathrm e}/T_{\mathrm p}$ in the interval $[0.1,10]$ show variations of the thresholds of $U_{\alpha}$ by less than one percent  for both instabilities under typical parameters. This confirms that the parallel-propagating drift/anisotropy instabilities are largely insensitive to electron temperatures within the range of possible solar-wind values.

\section{Contours of Constant Maximum Growth Rate}
\label{sec:contours} 

In this section, we present analytic fits to the 
contours of constant maximum growth rate $\gamma_{\mathrm
  m}$ in our numerical solutions to the hot-plasma dispersion relation.
In all cases, we have set $T_{\mathrm
  p}=T_{\mathrm e}=T_{\parallel \alpha}/4$ and
$T_{\perp}=T_{\parallel}$ for protons and electrons. As we have shown in Section~\ref{sect:density}, the instability thresholds are density-dependent in general. For the fits presented in this section, however, we set $n_{\alpha}=0.05n_{\mathrm p}$ as the typical observed value in the fast solar wind.

\subsection{Alfv\'en/Ion-Cyclotron (A/IC) Instability}

In the $U_{\alpha}/v_{\mathrm A}$-$w_{\parallel \alpha}/v_{\mathrm A}$ plane as in Figure~\ref{fig_ic-dati-Uw}, the  NHDS solutions show that two values of $U_{\alpha}/v_{\mathrm A}$ can potentially lead to the same maximum growth rate $\gamma_{\mathrm m}$ at one given value of $w_{\parallel \alpha}/v_{\mathrm A}$. For example, the points representing $\gamma_{\mathrm m}=10^{-4}\Omega_{\mathrm p}$ for $T_{\perp\alpha}/T_{\parallel \alpha}=1.2$ in Figure~\ref{fig_ic-dati-Uw} indicate that both $U_{\alpha}\approx v_{\mathrm A}$ and $U_{\alpha}\approx 0.6v_{\mathrm A}$ lead to the same maximum growth rate at $w_{\parallel\alpha}\approx 0.7v_{\mathrm A}$. Therefore, we employ two fits to describe the isocontours of constant $\gamma_{\mathrm m}$. In the range $0\le U_{\alpha}\le U_{\mathrm{med}}$, we apply a linear fit of the form
\begin{equation}\label{fit1}
\frac{w_{\parallel \alpha}}{v_{\mathrm A}}=a\frac{U_{\alpha}}{v_{\mathrm A}}+b.
\end{equation}
In the range $U_{\mathrm{med}}<U_{\alpha}\le 1.1v_{\mathrm A}$, we assume a shifted power-law function of the form
\begin{equation}\label{fit2}
\frac{w_{\parallel \alpha}}{v_{\mathrm A}}=A\left(\frac{U_{\alpha}}{v_{\mathrm A}}\right)^B+C.
\end{equation} 
The parameters for these fits are given in Table~\ref{tab_fits1}.

\begin{deluxetable*}{c|c|cc|ccc|c}
\tablecaption{Fit Parameters for Isocontours of Constant $\gamma_{\mathrm m}$ of the A/IC Instability in the $U_{\alpha}/v_{\mathrm A}$-$w_{\parallel \alpha}/v_{\mathrm A}$ Plane for Use in Equations~(\ref{fit1}) and (\ref{fit2})  \label{tab_fits1}}
\tablehead{ \colhead{$T_{\perp \alpha}/T_{\parallel \alpha}$} & \colhead{$\gamma_{\mathrm m}/\Omega_{\mathrm p}$} & \colhead{$a$} & \colhead{$b$} &  \colhead{$A$} & \colhead{$B$} & \colhead{$C$ } & \colhead{$U_{\mathrm{med}}/v_{\mathrm A}$ }   }
\startdata 
1.2 & $1\times 10^{-4}$ & -2.264 & 1.997 & 0.312 & 2.34 & 0.389 & 0.65 \\ 
1.2 & $5\times 10^{-4}$ & -2.510 & 2.366 & 0.174 & 4.41 & 0.750 & 0.65\\
1.2 & $1\times 10^{-3}$ & -2.576 & 2.599 & 0.164 & 3.38 & 0.952 & 0.65 \\
1.2 & $3\times 10^{-3}$ & -2.40 & 3.188 & 0.046 & 8.7 & 1.585 & 0.75 \\
1.2 & $1\times 10^{-2}$ & -3.28 & 5.807 & 0.305 & -2.10 & 3.093 & 0.65 \\  \hline
1.4 & $1\times 10^{-3}$ & -1.365 & 1.335 & 0.454 & 1.95 & 0.481 & 0.55 \\
1.4 & $3\times 10^{-3}$ & -1.328 & 1.661 & 0.350 & 2.55 & 0.916 & 0.55\\
1.4 & $1\times 10^{-2}$ & -1.118 & 2.622 & 0.146 & 4.39 & 2.053 & 0.55 \\ \hline
1.8 & $1\times 10^{-3}$ & -0.724 & 0.649 & 0.735 & 1.210 & 0.067 & 0.45 \\
1.8 & $3\times 10^{-3}$ & -0.699 & 0.813 & 0.726 & 1.346 & 0.286 & 0.45 \\
1.8 & $1\times 10^{-2}$ & -0.589 & 1.242 & 0.570 & 1.87 & 0.962 & 0.35 \\ \hline
2.0 & $1\times 10^{-4}$ & -0.471 & 0.371 & 0.602 & 1.225 & -0.038 & 0.35 \\
2.0 & $5\times 10^{-4}$ & -0.505 & 0.449 & 1.16 & 0.58 & -0.47 & 0.40 \\
2.0 & $1\times 10^{-3}$ & -0.583 & 0.507 & 0.801 & 1.080 & -0.041 & 0.40 \\
2.0 & $1\times 10^{-2}$ & -0.462 & 0.971 & 0.701 & 1.548 & 0.682 & 0.35 
\enddata
\end{deluxetable*}

We also give fit results for the A/IC instability in the $T_{\perp \alpha}/T_{\parallel \alpha}$-$w_{\parallel \alpha}/v_{\mathrm A}$ plane for different values of $U_{\alpha}/v_{\mathrm A}$ as shown in Figure~\ref{fig_dati_baleplot}. We use a similar fitting function to the formula suggested by \citet{hellinger06}:
\begin{equation}\label{fitfunct_baleplot}
\frac{T_{\perp \alpha}}{T_{\parallel \alpha}}=1+a\left(\frac{w_{\parallel \alpha}}{v_{\mathrm A}}-\frac{w_0}{v_{\mathrm A}}\right)^{-b}.
\end{equation}
The fitting parameters for a choice of values of $U_{\alpha}/v_{\mathrm A}$ and $\gamma_{\mathrm m}$ are given in Table~\ref{tab_fits2}. These parameters are valid in the range $w_{\min}\le w_{\parallel \alpha}\le 7v_{\mathrm A}$.

\begin{deluxetable}{c|c|ccc|c}
\tablecaption{Fit Parameters for Isocontours of Constant $\gamma_{\mathrm m}$ of the A/IC Instability in the $T_{\perp \alpha}/T_{\parallel \alpha}$-$w_{\parallel \alpha}/v_{\mathrm A}$ Plane for Use in Equation~(\ref{fitfunct_baleplot})  \label{tab_fits2}}
\tablehead{ \colhead{$U_{\alpha}/v_{\mathrm A}$} & \colhead{$\gamma_{\mathrm m}/\Omega_{\mathrm p}$} & \colhead{$a$} & \colhead{$b$} &  \colhead{$w_{0}/v_{\mathrm A}$} & \colhead{$w_{\min}/v_{\mathrm A}$}  }
\startdata 
0 & $10^{-4}$ & 0.394 & 0.958 & -0.0150 & 0.10 \\
0 & $10^{-3}$ & 0.529 & 0.949 & -0.0143 & 0.10 \\
0 & $10^{-2}$ & 0.977 & 0.876 & 0.0036 & 0.10 \\ \hline
0.4 & $10^{-4}$ & 0.219 & 0.92 & 0.000 & 0.15 \\
0.4 & $10^{-3}$ & 0.272 & 0.712 & 0.124 & 0.15 \\
0.4 & $10^{-2}$ & 0.749  & 0.853 & 0.134 & 0.20 
\enddata
\end{deluxetable}

\subsection{Fast-magnetosonic/Whistler (FM/W) Instability}

We first provide fit results for the FM/W instability in the $U_{\alpha}/v_{\mathrm A}$-$w_{\parallel \alpha}/v_{\mathrm A}$ plane as shown in Figure~\ref{fig_fh-dati-Uw}. The NHDS solutions are well represented by a parabolic function of the form
\begin{equation}\label{fitfunct_fhdatiUw}
\frac{U_{\alpha}}{v_{\mathrm A}}=a\left(\frac{w_{\parallel \alpha}}{v_{\mathrm A}}\right)^2+b\frac{w_{\parallel \alpha}}{v_{\mathrm A}}+c.
\end{equation}
The corresponding fitting parameters for different temperature anisotropies and growth rates are given in Table~\ref{tab_fits3}. The formulae are valid in the range $0\le U_{\alpha}\le 1.1v_{\mathrm A}$.
\begin{deluxetable}{c|c|ccc}
\tablecaption{Fit Parameters for Isocontours of Constant $\gamma_{\mathrm m}$ of the FM/W Instability in the $U_{\alpha}/v_{\mathrm A}$-$w_{\parallel \alpha}/v_{\mathrm A}$ Plane for Use in Equation~(\ref{fitfunct_fhdatiUw})  \label{tab_fits3}}
\tablehead{ \colhead{$T_{\perp \alpha}/T_{\parallel \alpha}$} & \colhead{$\gamma_{\mathrm m}/\Omega_{\mathrm p}$} & \colhead{$a$} & \colhead{$b$} &  \colhead{$c$}  }
\startdata 
0.8 & $5\times 10^{-5}$ & 0.047 & -0.689 & 1.408 \\
0.8 & $1\times 10^{-4}$ & 0.037 & -0.625 & 1.401 \\ 
0.8 & $1\times 10^{-3}$ & 0.000 & -0.359 & 1.352 \\
0.8 & $3\times 10^{-3}$ & -0.011 & -0.205 & 1.311 \\
0.8 & $1\times 10^{-2}$ & -0.0288 & 0.039 & 1.170 \\ \hline
0.5 & $1\times 10^{-3}$ & 0.198 & -1.440 & 1.789  \\
0.5 & $3\times 10^{-3}$ & 0.083 & -0.990 & 1.681 \\ 
0.5 & $1\times 10^{-2}$ & -0.0352 & -0.384 & 1.468 \\ \hline
0.4 & $5\times 10^{-5}$ & 0.570 & -2.717 & 2.019 \\
0.4 & $1\times 10^{-4}$ & 0.514 & -2.541 & 2.006\\
0.4 & $1 \times 10^{-3}$ & 0.315 & -1.826 & 1.923 \\
0.4 & $1\times 10^{-2}$ & -0.0144 & -0.568 & 1.560 
\enddata
\end{deluxetable}

We apply Equation~(\ref{fitfunct_baleplot}) also to the FM/W mode in order to describe isocontours of constant $\gamma_{\mathrm m}$ in the $T_{\perp \alpha}/T_{\parallel \alpha}$-$w_{\parallel \alpha}/v_{\mathrm A}$ plane. The resulting fit parameters are given in Table~\ref{tab_fits4}. These fits are valid in the range $w_{\min}\le w_{\parallel \alpha}\le 7v_{\mathrm A}$.

\begin{deluxetable}{c|c|ccc|c}
\tablecaption{Fit Parameters for Isocontours of Constant $\gamma_{\mathrm m}$ of the FM/W Instability in the $T_{\perp \alpha}/T_{\parallel \alpha}$-$w_{\parallel \alpha}/v_{\mathrm A}$ Plane for Use in Equation~(\ref{fitfunct_baleplot})  \label{tab_fits4}}
\tablehead{ \colhead{$U_{\alpha}/v_{\mathrm A}$} & \colhead{$\gamma_{\mathrm m}/\Omega_{\mathrm p}$} & \colhead{$a$} & \colhead{$b$} &  \colhead{$w_{0}/v_{\mathrm A}$} & \colhead{$w_{\min}/v_{\mathrm A}$}  }
\startdata 
0 & $10^{-4}$ & -0.345 & 0.742 & 0.530 & 0.80 \\
0 & $10^{-3}$ & -0.479 & 0.755 & 0.640 & 1.28 \\
0 & $10^{-2}$ & -1.052 & 0.900 & 0.697 & 1.90 \\ \hline
0.4 & $10^{-4}$ & -0.248 & 0.717 & 0.458 & 0.65 \\
0.4 & $10^{-3}$ & -0.347 & 0.707 & 0.556 & 0.84 \\
0.4 & $10^{-2}$ & -0.812  & 0.831 & 0.525 &  1.47
\enddata
\end{deluxetable}

\section{Quasilinear Evolution of the Alpha-Particle Distribution Function}\label{sect:ql}

Quasilinear theory is a theoretical framework for describing the evolution of a plasma under the effects of resonant wave--particle interactions. Prerequisites for the application of this description are small amplitudes and small growth or damping rates of the resonant waves. These assumptions imply that the background distribution function changes on a timescale that is much larger than the wave periods.
Resonant particles of species $s$ diffuse in velocity space according to the equation
\begin{multline}
\diffp{f_{s}}{t}=\lim _{V\to \infty}\sum \limits_{n=-\infty}^{+\infty}\frac{q_{s}^2}{8 \pi^2m_{s}^2}\int \frac{1}{Vv_{\perp}}\hat Gv_{\perp}\\
\times \delta(\omega_{k\mathrm r}-k_{\parallel}v_{\parallel}-n\Omega_{s}) \left|\psi_{n,k} \right|^2\hat Gf_{s}\mathrm d^3k\label{qldiff}
\end{multline}
\citep{stix92}.
The diffusive flux of particles in velocity space is always tangent to semicircles centered on the parallel phase velocity, which satisfy the equation
\begin{equation}\label{qlcircle}
\left(v_{\parallel}-v_{\mathrm{ph}}\right)^2+v_{\perp}^2=\mathrm{constant},
\end{equation} 
where $v_{\mathrm{ph}}\equiv \omega_{k\mathrm r}/k_{\parallel}$.
At the same time, Equation~(\ref{qldiff}) allows for diffusion only from higher phase-space densities to lower phase-space densities and, hence, resolves the ambiguity of the tangential direction given by Equation~(\ref{qlcircle}).

 Equations~(\ref{growthrate}) and (\ref{qldiff}) together fulfill energy conservation \citep{kennel67,chandran10b}. The alignment of the gradients in velocity space and the semicircles given by Equation~(\ref{qlcircle}) determine if the particles lose or gain kinetic energy during the quasilinear diffusion process. When the resonant particles lose kinetic energy, the wave gains energy and becomes unstable.

\begin{figure}
\epsscale{.8}
\plotone{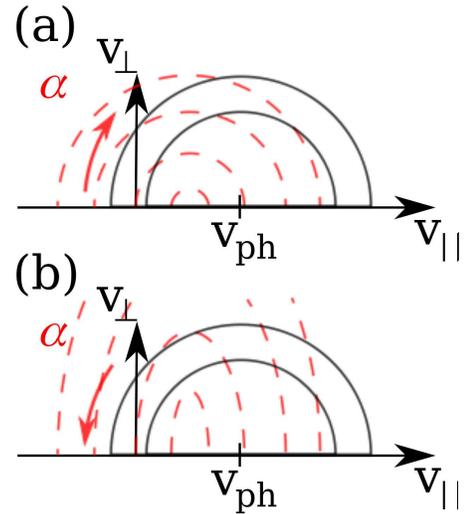}
\caption{Direction of the diffusive particle flux in velocity space for the resonance with $n=1$ (A/IC mode). Resonant particles diffuse in the direction tangent to semicircles (black solid lines) centered on the parallel phase velocity $v_{\mathrm{ph}}\equiv \omega_{k\mathrm r}/k_{\parallel}$ at $v_{\parallel}=v_{\mathrm{res}}<0$ from higher phase-space density to lower phase-space density as shown by the red arrow. The drift speed is less than the parallel phase speed of the waves. (a) Isotropic case with $T_{\perp \alpha}=T_{\parallel \alpha}$. The diffusing particles gain kinetic energy and, therefore, damp the wave. (b) Case with $T_{\perp \alpha}>T_{\parallel \alpha}$. The diffusing particles lose kinetic energy, thereby acting to destabilize the wave.}
\label{fig_diff_paths_icdati}
\end{figure}

Figure~\ref{fig_diff_paths_icdati} illustrates how particles lose energy by interacting with unstable A/IC waves. The resonant alpha particles that drive the A/IC instability typically have a parallel velocity $v_{\parallel}<0$ in order to be able to fulfill Equation~(\ref{rescond}), which is equivalent to $v_{\parallel}=v_{\mathrm{res}}$ from Equation~(\ref{vres}), with $n=1$.  Negative values of $v_{\parallel}$ (as measured in the proton rest frame) are needed for the instability of the A/IC mode when $\beta_{\mathrm p}\gtrsim 1$, because protons damp the waves that resonate with alphas that have $v_{\parallel}>0$. Figure~\ref{fig_diff_paths_icdati} shows how the interplay between temperature anisotropy and relative drift leads to a change in the alignment of the gradients of $f_{\alpha}$ in velocity space and, therefore, to a transition from a damped to an unstable situation.

\begin{figure}
\epsscale{.8}
\plotone{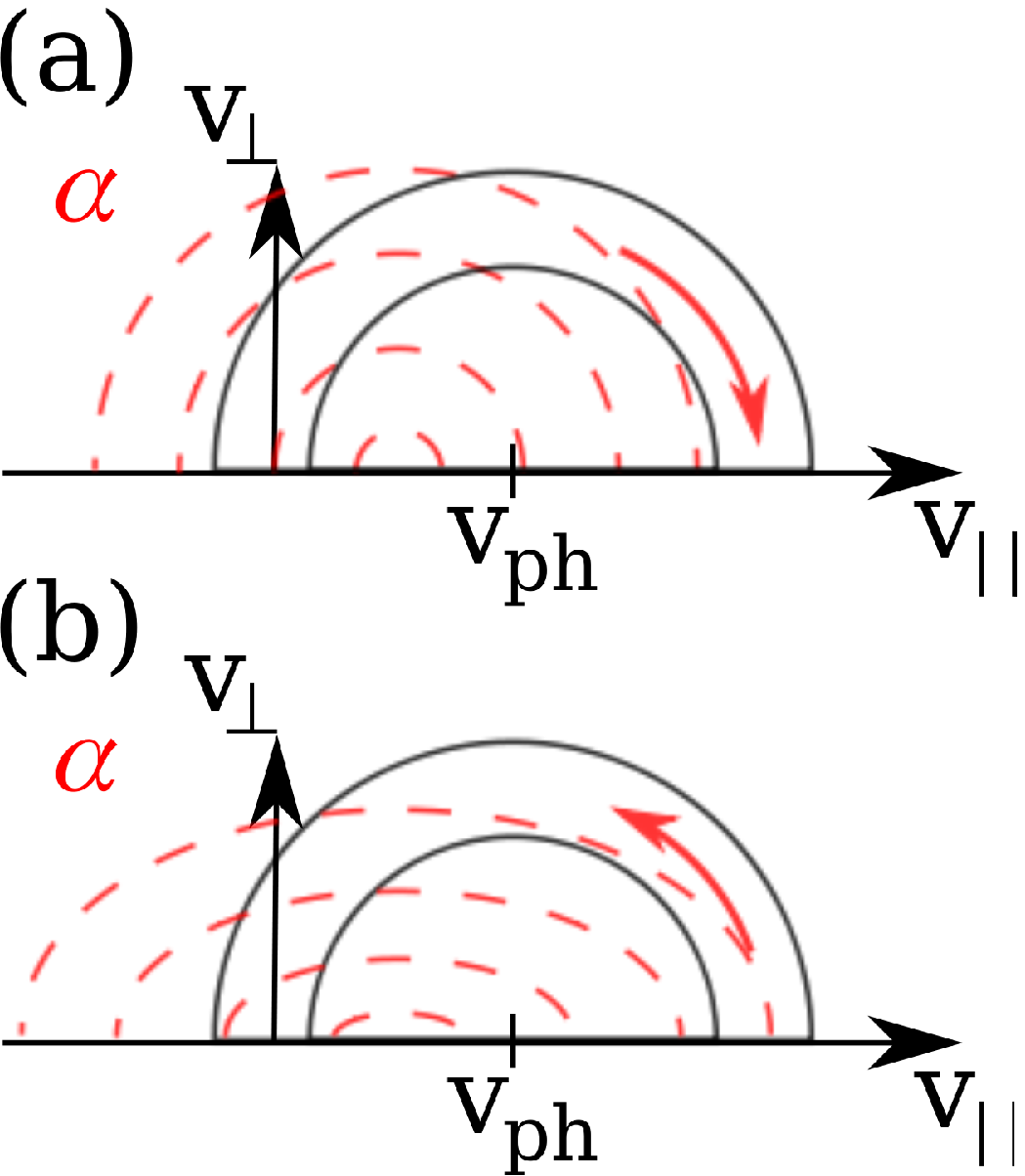}
\caption{Direction of the diffusive particle flux in velocity space for the resonance with $n=-1$ (FM/W mode). Resonant particles diffuse in the direction tangent to semicircles (black solid lines) centered on the parallel phase velocity $v_{\mathrm{ph}}\equiv \omega_{k\mathrm r}/k_{\parallel}$ at $v_{\parallel}=v_{\mathrm{res}}>0$ from higher phase-space density to lower phase-space density as shown by the red arrow. The drift speed is less than the parallel phase speed of the waves. (a) Isotropic case with $T_{\perp \alpha}=T_{\parallel \alpha}$. The diffusing particles gain kinetic energy and, therefore, damp the wave. (b) Case with $T_{\perp \alpha}<T_{\parallel \alpha}$. The diffusing particles lose kinetic energy, thereby acting to destabilize the wave.}
\label{fig_diff_paths_fhdati}
\end{figure}

The situation for the FM/W instability in velocity space is illustrated in Figure~\ref{fig_diff_paths_fhdati}. This figure shows how particles with $v_{\parallel}>\omega_{k\mathrm r}/k_{\parallel}>U_{\alpha}$ can lose or gain kinetic energy depending on the alignment between velocity-space gradients and the diffusion paths.

Quasilinear diffusion reduces the temperature anisotropy and the relative drift, which are the two sources of free energy for the A/IC and FM/W instabilities, as illustrated by the initial diffusion paths in Figures~\ref{fig_diff_paths_icdati} and \ref{fig_diff_paths_fhdati}. Since only part of the distribution function participates in the diffusion, the resulting distribution function can not be described as a more isotropic bi-Maxwellian at lower drift speed, but will show additional non-Maxwellian structures. Such a shaped distribution function will carry waves with modified dispersion relations compared to the initially bi-Maxwellian distribution function. The treatment of such waves, however, lies beyond the scope of this paper.

\section{Regulation of $U_{\alpha}$ by instabilities: the Possible Role of Large-Scale Magnetic-Field-Strength Fluctuations}\label{sect:ani_U}

\begin{figure}
\epsscale{1.2}
\plotone{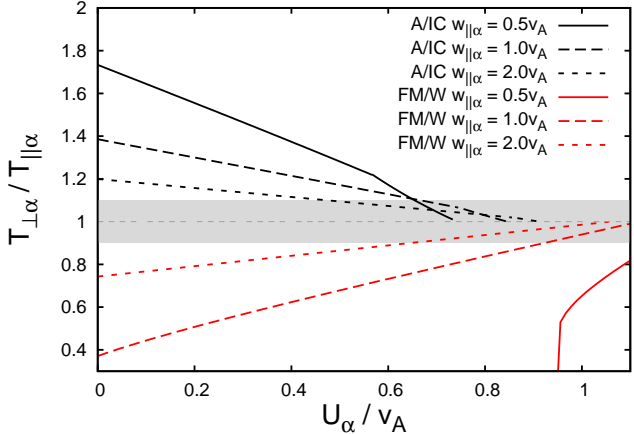}
\caption{Instability thresholds in the $T_{\perp \alpha}/T_{\parallel \alpha}$-$U_{\parallel \alpha}/v_{\mathrm A}$ plane for different values of $w_{\parallel \alpha}/v_{\mathrm A}$. We show our analytical thresholds for the A/IC mode with the more restrictive criterion from Equations~(\ref{cond1ani}) or (\ref{eq60ani}), respectively, and the analytical threshold for the FM/W modes from (\ref{cond2}). We set $\sigma=2.4$ for the A/IC mode and $\sigma=2.1$ for the FM/W mode. The grey shaded box indicates the range $0.9\le T_{\perp \alpha}/T_{\parallel \alpha}\le 1.1$.}
\label{fig_ani_U}
\end{figure}

In Figure~\ref{fig_ani_U}, we show the unstable parameter space in the plane $T_{\perp \alpha}/T_{\parallel \alpha}$ versus $U_{\alpha}/v_{\mathrm A}$ for three different values of $w_{\parallel \alpha}/v_{\mathrm A}$.
For each value of $w_{\parallel \alpha}/v_{\mathrm A}$, the unstable region lies above and to the right of the A/IC curves, and below and to the right of the FM/W curves.
Variations in $|\vec B|$ at around the 10\% level are observed in the
solar wind \citep{tu95}. As alpha particles flow away from the Sun (and past the protons),
they alternately move through regions of larger $|\vec B|$ and smaller
$|\vec B|$, which causes $T_{\perp \alpha}/T_{\parallel \alpha}$ to
fluctuate. A qualitative argument that supports this claim follows from the idea of double-adiabatic expansion: if alpha-particle heating and the alpha-particle heat flux are neglected, then $T_{\perp\alpha}/T_{\parallel \alpha}\propto B^3/n_{\alpha}^2$ \citep{chew56,sharma06a}. As $T_{\perp \alpha}/T_{\parallel \alpha}$ varies, the
instability threshold varies. For some solar-wind streams, the alphas
may alternate between being susceptible to the A/IC instability and being
susceptible to the FM/W instability. For other streams, the sign of
$T_{\perp \alpha}/T_{\parallel \alpha}-1$ may remain fixed, but the
magnitude of this quantity will vary. If at any point, the
distribution becomes unstable, the alphas will decelerate (and
isotropize), and the deceleration will be cumulative. As a result, due
to the combined action of $|\vec B|$ variations and scattering from
instabilities, the alphas can evolve to values of $U_{\alpha}$ and
$T_{\perp \alpha}/T_{\parallel \alpha}$ that are below the local
instability threshold \citep[see also][]{seough13}.
This process is illustated by the shaded rectangle in Figure~\ref{fig_ani_U} that represents the range $0.9\le T_{\perp \alpha}/T_{\parallel \alpha}\le 1.1$. For a thermal speed of $w_{\parallel \alpha}=2v_{\mathrm A}$, this effect drives the plasma alternately into the A/IC instability and into the FM/W instability for $U_{\alpha}\gtrsim0.6v_{\mathrm A}$ if $T_{\perp \alpha}/T_{\parallel \alpha}$ varies in that range. In this way, even a plasma that is isotropic on average can be limited in $U_{\alpha}$ to a significantly lower value than prescribed by the thresholds of the isotropic beam instabilities.

\section{Conclusions}\label{sec:conc}

We derive analytic instability criteria for parallel-propagating
($k_\perp=0$) A/IC waves and
FM/W waves in the presence of alpha
particles that drift with respect to protons at an average
velocity~$U_{\alpha} \vec B_0/B_0$.  We take the alpha particles to
have a bi-Maxwellian distribution function and the protons and
electrons to be Maxwellian, and we focus on the case in which $w_{\parallel \mathrm p}\gtrsim 0.25v_{\mathrm A}$. Our analysis is
based on \citeauthor{kennel67}'s \citeyearpar{kennel67} expression for the growth
rate~$\gamma_{k}$ of plasma waves satisfying the inequality $|\gamma_k|
\ll |\omega_{k\mathrm r}|$.  Rather than evaluate the full expression for
$\gamma_k$, we undertake the more modest task of evaluating the sign
of $\gamma_k$. \citeauthor{kennel67}'s \citeyearpar{kennel67} formula is rigorously correct
in the limit $|\gamma_k| \ll |\omega_{k\mathrm r}|$, but requires knowledge of
the precise value of the real part of the frequency~$\omega_{k\mathrm r}$.  Our
analysis is based on an approximation to the dispersion relation, as
well as an approximate treatment of proton cyclotron damping, and our
results are therefore not exact. However, as we summarize further
below, our approximate analytic results are in good agreement with
numerical solutions to the full hot-plasma dispersion relation, which
gives us confidence that our approximations are reasonable.

We find that the left-circularly polarized A/IC wave is driven
unstable by the $n=1$ resonance (see Equation~(\ref{rescond})) with alpha
particles that satisfy $v_\parallel < 0$ in the proton frame.  In the non-drifting limit ($U_{\alpha}=0$), this instability corresponds to the parallel ion-cyclotron instability \citep[e.g.,][]{sagdeev61,gary94}. In the isotropic limit ($T_{\perp \alpha}=T_{\parallel \alpha}$), this instability corresponds to the parallel Alfv\'enic instability described by \citet{verscharen13a}. There
are two separate instability criteria for this mode.  First, the
alpha-particle parallel thermal speed $w_{\parallel \alpha}$ needs to
be sufficiently large that there is a wavenumber interval~$[k_1, k_2]$
in which thermal alpha particles can resonate with the waves but
thermal protons cannot. Within this interval, proton cyclotron damping
can be neglected. Second, the alpha-particle drift velocity $U_\alpha$
and/or temperature anisotropy $T_{\perp \alpha}/T_{\parallel \alpha}$
need to be sufficiently large that resonant interactions between alpha
particles and A/IC waves are destabilizing in the interval~$[k_1,
k_2]$. The mathematical expressions of these criteria are given in
Section~\ref{subsect:aic}.

The right-circularly polarized FM/W wave is driven unstable by the
$n=-1$ resonance with alpha particles with $v_\parallel > 0$ in the
proton frame.  In the non-drifting limit ($U_\alpha=0$), this
instability corresponds to the parallel firehose instability \citep{quest96,rosin11}. In the isotropic limit ($T_{\perp \alpha}=T_{\parallel \alpha}$), this instability corresponds to the
parallel magnetosonic instability \citep{gary00b,li00}.  
The FM/W wave is unstable if and only if the following conditions are satisfied. First, $U_{\alpha}+\sigma w_{\parallel \alpha}$ needs to be sufficiently large that thermal alpha particles can resonate with the FM/W wave. Second, $U_{\alpha}$ and/or  $T_{\parallel \alpha}/T_{\perp \alpha}$ need to be sufficiently
large that there is an interval of wavenumbers, say $[k_3,k_4]$, within which resonant interactions with alpha particles are destabilizing. Third, $w_{\parallel \mathrm p}$ must be less than a certain threshold (Equation~(\ref{protdamp})), so that resonant protons are unable to damp the FM/W waves within at least part of the wavenumber interval $[k_3,k_4]$. The mathematical expressions of these criteria are given in Section~\ref{sect:DFI}.

The quantity $\sigma$ that appears in our
analytic instability criteria is the maximum number of
thermal speeds that the resonant particles' $v_\parallel$ can be from
the center of the distribution before there are too few resonant
particles to cause significant amplification or damping of a mode. For
example, if the only alpha particles that can resonate with a
particular wave are ten thermal speeds out on the tail of the
distribution, then there are not enough resonant alpha particles to
lead to an appreciable growth rate for that wave. In our analytic
calculations, $\sigma$ functions like a free parameter and is not
determined from first principles. The best choice for the value of
$\sigma$ depends upon what is meant by ``significant amplification''
or ``appreciable growth rate'' in the above discussion.  We choose to
set $\sigma = 2.4$ for the A/IC instability and $\sigma=2.1$ for the FM/W instability for plasmas with $n_{\alpha} =0.05 n_{\mathrm
  p}$. Using numerical solutions to the full hot-plasma dispersion
relation, we find that this choice causes our analytic instability
criteria to correspond to parameter combinations for which the maximum
growth rate is $\simeq 10^{-4} \Omega_{\mathrm p}$ for A/IC waves and
 FM/W waves. In Section~\ref{sect:density},
we argue that this choice of $\sigma$ can be generalized to other
solar-wind-relevant values of $n_{\alpha}/n_{\mathrm p}$ by taking $\sigma= [ \ln (6250 n_\alpha/n_{\mathrm p})]^{1/2}$ for the alpha particles and $\sigma=2.4$ for the protons in the case of the A/IC instability, and by taking $\sigma= [ \ln (1640 n_\alpha/n_{\mathrm p})]^{1/2}$ for the alpha particles and $\sigma=2.1$ for the protons in the case of the FM/W instability. We present numerical solutions
to the hot-plasma dispersion relation to support this assertion.  With
this generalization, our analytic instability criteria can be used to
determine whether A/IC and FM/W waves are stable or unstable as a
function of the following five dimensionless parameters:
$w_{\parallel \alpha}/v_{\mathrm A}$, $U_{\alpha}/v_{\mathrm A}$, $T_{\perp \alpha}/T_{\parallel \alpha}$, $T_{\parallel \alpha}/T_{\mathrm p}$, and
$n_{\alpha}/n_{\mathrm p}$.  We also
present arguments and numerical evidence in Section~\ref{sect:density} that the
instability thresholds of the parallel-propagating A/IC and FM/W waves
are largely insensitive to the value of a sixth parameter, $T_{\mathrm
  e}/T_{\mathrm p}$. Knowledge of how these instability thresholds depend
upon the above parameters will be useful for comparing these
thresholds with spacecraft observations to test the extent to which
A/IC and FM/W instabilities limit the differential flow and
temperature anisotropy of alpha particles in the solar wind.

In addition to deriving analytic expressions for the instability
thresholds, we use numerical solutions to the hot-plasma dispersion relation to
find analytic fits to contours of constant maximum growth rates for
A/IC and FM/W waves in both the $T_{\perp \alpha}/T_{\parallel \alpha}$-$w_{\parallel \alpha}/v_{\mathrm A}$ and $U_{\alpha}/v_{\mathrm A}$-$w_{\parallel \alpha}/v_{\mathrm
  A}$ planes. These fits are presented in
Section~\ref{sec:contours}.  In Section~\ref{sect:ql}, we use quasilinear theory to describe how some of the instability criteria can be understood in terms of the
change in the alpha particles' kinetic energy during resonant
wave--particle interactions.

Finally, in Section~\ref{sect:ani_U} we discuss the possible role that large-scale
magnetic-field fluctuations play in the regulation of $U_{\alpha}$ and
$T_{\perp \alpha}/T_{\parallel \alpha}$. Spacecraft measurements
indicate that the magnetic field strength in the solar wind varies on
roughly hour-long time scales at the $\sim 10\%$ level~\citep{tu95}. As
alpha particles alternately move through larger-$|\vec B|$ and
smaller-$|\vec B|$ regions, the value of $T_{\perp \alpha}/T_{\parallel \alpha}$ fluctuates, even in the absence of
heating~\citep{chew56,sharma06a}. These fluctuations in
$T_{\perp \alpha}/T_{\parallel \alpha}$ subject the alpha particles to a
time-varying $U_\alpha/v_{\mathrm A}$ instability threshold. As variations
in~$|\vec B|$ cause alpha particles to evolve from maximum anisotropy
towards isotropy, the $U_{\alpha}/v_{\mathrm A}$ instability threshold for
these alpha particles increases, but $U_{\alpha}$ does not, so that
the alpha particles can evolve away from the instability thresholds
and into the stable part of parameter space, even if collisions are
negligible.

One of the limitations of our analysis is that we have taken the
alpha-particle distribution function to be bi-Maxwellian in order to simplify the analysis. In the solar
wind, the actual distribution function is probably never bi-Maxwellian
\citep[e.g.,][]{isenberg12}, and thus our treatment is only an
approximation. Two other limitations of our analysis are that we have
neglected proton temperature anisotropy and restricted our analysis to
the case of parallel-propagating waves with $k_\perp = 0$. It will be
important to relax these limitations in future research in order to
further advance our understanding of kinetic plasma instabilities in
the solar wind.

\acknowledgements

We appreciate helpful discussions with Ben Maruca and Alex Schekochihin. This work was supported in part by grant NNX11AJ37G
from NASA's Heliophysics Theory Program, NASA grant NNX12AB27G,
NSF/DOE grant AGS-1003451, and DOE grant DE-FG02-07-ER46372.

\appendix
\section{Comparing the Two Instability Thresholds for the A/IC Wave}

In Section~\ref{subsect:aic}, we found that the two necessary and sufficient
conditions for an instability of the A/IC wave are, first, Equation~(\ref{cond3b})
and, second, either Equations~(\ref{cond1b}) and (\ref{cond2b}) or Equation~(\ref{eq60}). In this appendix, we describe the conditions under which Equations~(\ref{cond1b}) and
(\ref{cond2b}) lead to the less restrictive ``second condition'' on
$U_{\alpha}$ and the conditions under which Equation~(\ref{eq60}) leads to the
less restrictive second condition on~$U_\alpha$.  Because it is only
necessary to satisfy either Equations~(\ref{cond1b}) and (\ref{cond2b}) or Equation~(\ref{eq60}),
it is the less restrictive of these two possibilities that is relevant
for determining the instability threshold.

\begin{figure}
\epsscale{1.2}
\plotone{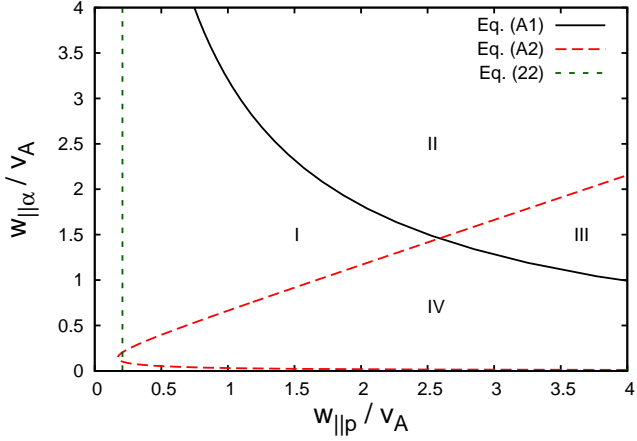}
\caption{Four regions of parameter space when $T_{\perp \alpha}=1.02T_{\parallel \alpha}$ and $\sigma=2.4$. In region I, the necessary and sufficient conditions for instability of the A/IC wave are Equations~(\ref{eq60}) and (\ref{upperlimitDCI}). In region II, the necessary and sufficient conditions for instability of the A/IC wave are Equations~(\ref{cond1b}), (\ref{cond2b}), and (\ref{upperlimitDCI}). No instability is possible in regions III and IV. }
\label{fig_test60_quad_low}
\end{figure}

After some algebra, we find that the right-hand sides of Equations~(\ref{cond1b}) and (\ref{eq60}) are equal if either of the following two equations is satisfied:\footnote{Technically,  Equation~(\ref{equalline}) implies that the right-hand sides of Equations~(\ref{cond1b}) and (\ref{eq60}) are equal only if the additional criterion  $\sigma w_{\parallel \mathrm p}/v_{\mathrm A} \ge \sqrt{2} - 1$ is also satisfied, but this condition is automatically satisfied because Equation~(\ref{k3defapp}) has already restricted our analysis to $\sigma w_{\parallel \mathrm p} > 0.5v_{\mathrm A}$.}
\begin{equation}\label{equalline}
\sigma w_{\parallel \mathrm p}=\left(\frac{T_{\perp \alpha}}{T_{\parallel \alpha}}-1\right)\sigma w_{\parallel \alpha}+\frac{T_{\parallel \alpha}}{2\sigma w_{\parallel \alpha}(T_{\perp \alpha}-T_{\parallel \alpha})}-v_{\mathrm A}
\end{equation}
or
\begin{equation}\label{equalline2}
\sigma w_{\parallel \mathrm p}=2\frac{T_{\perp \alpha}}{T_{\parallel \alpha}}\sigma w_{\parallel \alpha}+\frac{v_{\mathrm A}^2T_{\parallel \alpha}}{4\sigma w_{\parallel \alpha}T_{\perp \alpha}}-v_{\mathrm A}.
\end{equation}
For a fixed value of $T_{\perp \alpha}/T_{\parallel \alpha}$, Equations~(\ref{equalline}) and (\ref{equalline2}) correspond to two curves in the
$w_{\parallel \alpha}/v_{\mathrm A}$-$w_{\parallel \mathrm p}/v_{\mathrm A}$ plane, which are plotted in Figure~\ref{fig_test60_quad_low} for the case in which $T_{\perp \alpha} = 1.02 T_{\parallel \alpha}$ and $\sigma = 2.4$.  
These two curves divide the
$w_{\parallel \alpha}/v_{\mathrm A}$-$w_{\parallel \mathrm p}/v_{\mathrm A}$ plane up into the four regions, which
are labeled in Figure~\ref{fig_test60_quad_low}. After some additional algebra, one can show
that the right-hand side of Equation~(\ref{upperlimitDCI}) (which is a restatement of
the instability condition in Equation~(\ref{cond3b})) is less than the
right-hand sides of Equations~(\ref{cond1b}) and (\ref{eq60}) in regions III and IV.
There is thus no interval of $U_\alpha$ values that can satisfy the
instability criteria in regions III and~IV.  On the other hand, the
right-hand side of Equation~(\ref{upperlimitDCI}) is greater than the right-hand sides
of Equations~(\ref{cond1b}) and (\ref{eq60}) in regions I and II, and an instability
is possible in these two regions. One can show that the
right-hand side of Equation~(\ref{eq60}) is smaller than the right-hand side
of Equation~(\ref{cond1b}) in region I, so that Equations~(\ref{cond3b}) and (\ref{eq60}) are the necessary and sufficient criteria for instability in region I. In
contrast, the right-hand side of Equation~(\ref{cond1b}) is smaller than the
right-hand side of Equation~(\ref{eq60}) in region II, and therefore Equations~(\ref{cond3b}) and (\ref{cond1b}) are the necessary and sufficient conditions for
instability in region II. The minimum value of
$\sigma w_{\parallel \alpha}$ in region~II is $(0.5v_{\mathrm
  A}T_{\parallel \alpha}/T_{\perp \alpha})[T_{\perp \alpha}/(T_{\perp \alpha} - T_{\parallel \alpha})]^{1/2}$, which is larger than the
lower limit in Equation~(\ref{cond2b}); Equation~(\ref{cond2b}) can thus be omitted as a
separate condition in region II.

\begin{figure}
\epsscale{1.2}
\plotone{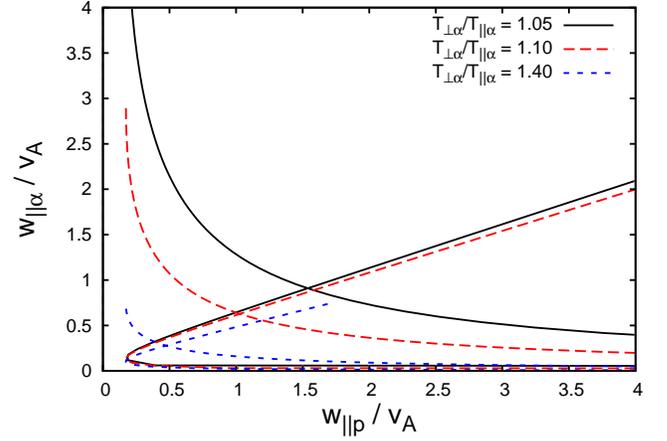}
\caption{Regions of parameter space in the $w_{\parallel \alpha}/v_{\mathrm A}$-$w_{\parallel \mathrm p}/v_{\mathrm A}$ plane for different $T_{\perp \alpha}/T_{\parallel \alpha}$ and $\sigma=2.4$. The hyperbolic curves opening upward and to the right are plots of Equation~(\ref{equalline}). The hyperbolic curves opening to the right with small opening angle are plots of Equation~(\ref{equalline2}).}
\label{fig_test60_quad_anis}
\end{figure}

In Figure~\ref{fig_test60_quad_anis}, we re-plot the lines from Equations~(\ref{equalline}) and (\ref{equalline2}) as in Figure~\ref{fig_test60_quad_low} but for different
 temperature anisotropies. The comparison between these two
figures illustrates how region~I contracts as the
temperature anisotropy increases.  The typical parameters of weakly
collisional solar-wind streams near 1~AU ($w_{\parallel \alpha} \simeq
w_{\parallel \mathrm p} \simeq v_{\mathrm A}$) correspond to region II when
$T_{\perp \alpha}/T_{\parallel \alpha} \gtrsim 1.1$ (so that the
instability criteria are Equations~(\ref{cond3b})-(\ref{cond2b})) and to region I when $T_{\perp \alpha}/T_{\parallel \alpha} \lesssim 1.1$ (so that the
instability criteria are Equations~(\ref{cond3b}) and (\ref{eq60})).

\begin{figure}
\epsscale{1.2}
\plotone{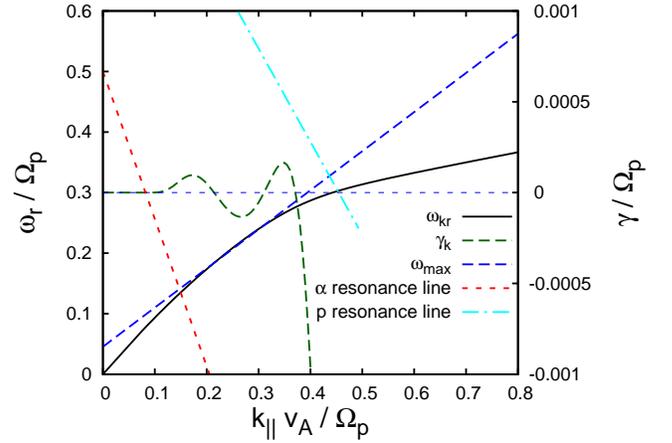}
\caption{Dispersion relation and resonance conditions for a plasma with $w_{\parallel \alpha}=2w_{\parallel \mathrm p}=1.4v_{\mathrm A}$, $T_{\perp \alpha}=1.1T_{\parallel \alpha}$, $T_{\mathrm p}=T_{\mathrm e}$, and $U_{\alpha}=0.646v_{\mathrm A}$. The frequency $\omega_{k\mathrm r}$ and growth rate $\gamma_k$ are solutions of the hot-plasma dispersion relation obtained from NHDS. This figure illustrates the interplay between the different instability and resonance conditions.}
\label{fig_test60_dispersiona}
\end{figure}

As mentioned in Section~\ref{subsect:aic}, Equation~(\ref{cond1b}) is the condition that
$\omega_{k\mathrm r} < \omega_{\max}^{\mathrm{A/IC}}$ at $k_\parallel= k_1$, and
Equation~(\ref{eq60}) is the condition that $\omega_{k\mathrm r} < \omega_{\max}^{\mathrm{A/IC}}$ at $k_\parallel= k_2$ (see Figure~\ref{fig_ic-dati} for the
definitions of these quantities). The existence of two alternative
conditions, Equations~(\ref{cond1b}) and (\ref{eq60}), thus rests on the idea that the
most unstable wavenumber can be located at either end of the interval
$[k_1, k_2]$. In Figure~\ref{fig_test60_dispersiona} we illustrate this possibility directly
using numerical solutions to the hot-plasma dispersion relation
obtained with the NHDS code~\citep{verscharen13a}. 
We plot the real and imaginary parts of the A/IC wave frequency for the case in which
$w_{\parallel \alpha} = 2w_{\parallel \mathrm p}$ and
$T_{\perp \alpha}=1.1T_{\parallel \alpha}$. The drift speed ($U_\alpha
= 0.646 v_{\mathrm A}$) is chosen in such a way that the maximum-frequency
line $\omega_{\mathrm r} = \omega_{\max}^{\mathrm{A/IC}}$ from Equation~(\ref{omegamax1}) intersects with the dispersion relation $\omega_{\mathrm r} = \omega_{k\mathrm r}$ twice
between $k_1$ and $k_2$. The wavenumber range in this diagram can be divided
into five intervals. In the first interval $(0 < k_\parallel v_{\mathrm
  A}/\Omega_{\mathrm p} < 0.15)$, thermal particles cannot resonate with
the waves, and $\gamma_k $ is vanishingly small.  In the second range
$(0.15 < k_\parallel v_{\mathrm A}/\Omega_{\mathrm p} < 0.22)$, which
contains~$k_1$, thermal alpha particles can resonate with the waves,
thermal protons cannot resonate, $\omega_{k\mathrm r} < \omega_{\max}^{\mathrm{A/IC}}$, and $\gamma_k > 0$.  In the next interval $(0.22 <
k_\parallel v_{\mathrm A}/\Omega_{\mathrm p} < 0.3)$, $\omega_{k\mathrm r} >
\omega_{\max}^{\mathrm{A/IC}}$, alpha particles damp the waves, and
$\gamma_k < 0$.  The growth rate then becomes positive again in the
fourth interval, $(0.3 < k_\parallel v_{\mathrm A}/\Omega_{\mathrm p} <
0.38)$, because $\omega_{k\mathrm r}$ is again~$< \omega_{\max}^{\mathrm{A/IC}}$. This fourth interval contains wavenumbers near~$k_2$. In the
fifth wavenumber interval $ (k_\parallel v_{\mathrm A}/\Omega_{\mathrm p} >
0.38)$, the number of resonant protons is sufficiently large to damp
the waves, and the damping rate increases with~$k_{\parallel}$ because of the
increase in the number of resonant protons.

\bibliographystyle{apj}
\bibliography{DATIs}

\end{document}